\documentclass[twocolumn,prd,floatfix,preprintnumbers,a4paper,nofootinbib,superscriptaddress]{revtex4-1}
\usepackage{epsfig}
\usepackage{graphics}
\usepackage{graphicx}
\usepackage{amsmath,amssymb}
\usepackage{amsfonts}
\usepackage{bm}
\usepackage[usenames,dvipsnames]{color}
\usepackage{amssymb}
\usepackage{psfrag}
\usepackage{times}
\usepackage[varg]{txfonts}
\usepackage{xspace}
\usepackage{float}
\usepackage{placeins} 
\usepackage{capt-of} 
\usepackage[colorlinks, pdfborder={0 0 0}]{hyperref} 
\usepackage[nameinlink,capitalise]{cleveref}

\definecolor{LinkColor}{rgb}{0.75, 0, 0}
\definecolor{CiteColor}{rgb}{0, 0.5, 0.5}
\definecolor{UrlColor}{rgb}{0, 0, 0.75}
\hypersetup{linkcolor=LinkColor}
\hypersetup{citecolor=CiteColor}
\hypersetup{urlcolor=UrlColor}

\let\oldtheequation\theequation
\makeatletter
\def\tagform@#1{\maketag@@@{\ignorespaces#1\unskip\@@italiccorr}}
\renewcommand{\theequation}{(\oldtheequation)}
\makeatother

\usepackage[utf8]{inputenc}
\usepackage{ulem}
\newcommand{\comment}[0]{\textcolor{black}}

\newcommand{\bns}[1][]{binary neutron star#1 (BNS#1)\renewcommand{\bns}[1][]{BNS##1\xspace}\xspace}
\newcommand{\nss}[1][]{Neutron stars#1 (NSs#1)\renewcommand{\nss}[1][]{NSs##1\xspace}\xspace}
\newcommand{\ns}[1][]{Neutron star#1 (NS#1)\renewcommand{\ns}[1][]{NS##1\xspace}\xspace}
\newcommand{\eos}[1][]{equation of state#1 (EOS#1)\renewcommand{\eos}[1][]{EOS##1\xspace}\xspace}
\newcommand{\snr}[1][]{signal-to-noise ratio#1 (SNR#1)\renewcommand{\snr}[1][]{SNR##1\xspace}\xspace}

\newcommand{\aLIGOe}{\texttt{eaLIGO}\xspace}

\newcommand{\ET}{\texttt{ET-D}\xspace}

\begin{document}


\title{Impact of high-order tidal terms on binary neutron-star waveforms}

\newcommand{\napoli}{\affiliation{Sezione INFN Napoli, Complesso Universitario di Monte S. Angelo, Via Cinthia, I-80126, 
Napoli, Italy}}
\newcommand{\sapienza}{\affiliation{Dipartimento di Fisica, ``Sapienza'' Universit\`a di Roma, Piazzale Aldo Moro 5, 00185, Roma, Italy}}
\newcommand{\infnroma}{\affiliation{Sezione INFN Roma1, Piazzale Aldo Moro 5, 00185, Roma, Italy}}

\author{Xisco Jim\'enez Forteza}\email{fjimenez@na.infn.it }\napoli\sapienza
\author{Tiziano Abdelsalhin}\email{tiziano.abdelsalhin@roma1.infn.it}\sapienza\infnroma
\author{Paolo Pani}\email{paolo.pani@roma1.infn.it}\sapienza\infnroma
\author{Leonardo Gualtieri}\email{leonardo.gualtieri@roma1.infn.it}\sapienza\infnroma

\begin{abstract}
GW170817, the milestone gravitational-wave event originated from a binary neutron star merger, has allowed scientific community to place a constraint on the equation of state of neutron stars by extracting the leading-order, tidal-deformability term from the gravitational waveform. Here we incorporate tidal corrections to the gravitational-wave phase at next-to-leading and next-to-next-to-leading order, including the magnetic tidal Love numbers, tail effects, and the spin-tidal couplings recently computed in Tiziano Abdelsalhin \textit{et al.}
[\href{https://journals.aps.org/prd/abstract/10.1103/PhysRevD.98.104046}{Phys. Rev. D 98, 104046 (2018)}]. These effects have not yet been included in the waveform approximants for the analysis of GW170817. We provide a qualitative and quantitative analysis of the impact of these new terms by studying the parameter bias induced on events compatible with GW170817 assuming second-generation (advanced LIGO) and third-generation (Einstein Telescope) ground-based gravitational-wave interferometers. We find that including the tidal-tail term deteriorates the convergence properties of the post-Newtonian expansion in the relevant frequency range. We also find that the effect of magnetic tidal Love numbers could be measurable for an optimal GW170817 event with signal-to-noise ratio $\rho \approx 1750$ detected with the Einstein Telescope. On the same line, spin-tidal couplings may be relevant if mildly high-spin $\chi \gtrsim 0.1$ neutron star binaries exist in nature.

\vspace{\baselineskip}
	\centering
	           published journal version: \\[0.5\baselineskip]
	           Phsyical Review D \textbf{98}, 124014 (2018) \\[0.5\baselineskip]
	           \href{https://doi.org/10.1103/PhysRevD.98.124014}{doi:10.1103/PhysRevD.98.124014}\\
	           \vspace{\baselineskip}
\end{abstract}

\maketitle

\section{Introduction}\label{sec:intro}

In August 2017, the LIGO-Virgo Scientific Collaboration reported the milestone detection of gravitational waves~(GWs)
from a \bns coalescence~\cite{TheLIGOScientific:2017qsa}, dubbed GW170817. This landmark discovery has opened a new era
in astrophysics. Along with the GW detection, several telescopes also reported the observation of electromagnetic
coincidence signals in various bands, inaugurating the birth of the GW multimessenger astronomy.

One of the most relevant implications of this discovery is arguably the possibility of constraining the equation of
state~(EoS) of the neutron-star (NS) core through the measurement of the tidal deformability of the binary
components~\cite{DelPozzo:2013ala,TheLIGOScientific:2017qsa,
  Bauswein:2017vtn,Most:2018hfd,Harry:2018hke,Abbott:2018exr,De:2018uhw, Abbott:2018wiz}.
GW170817 allowed one to place stringent constraints on the leading-order, tidal-deformability parameter, which measures the
induced quadrupole moment of the binary components due to tidal forces during the late inspiral.
Though in the past some constraints on the radius of isolated NSs (and hence on the EoS) have been set based on
electromagnetic observations (see Ref.~\cite{Ozel:2016oaf} for a review), the GW channel is expected to provide more
robust and tighter constraints, especially as more events are observed by current and next-generation GW detectors.

GW searches and parameter estimation pipelines rely on waveform approximants that accurately describe the inspiral,
merger, and ringdown phases of the coalescence. While the early inspiral is accurately described by the
post-Newtonian~(PN) theory~\cite{Arun:2008kb,Buonanno:2009zt,Mishra:2016whh}, this description breaks down near the
merger. To overcome this limitation, current waveform templates are recalibrated by fitting the late inspiral and merger
phase to numerical relativity~(NR) solutions~\cite{Dietrich:2017aum,Dietrich:2017feu,Dietrich:2018uni,Cotesta:2018fcv},
producing the so-called Phenom~\cite{Santamaria:2010yb,Hannam:2013oca,Husa:2015iqa,Khan:2015jqa} and SEOBNR
approximants~\cite{Pan:2013rra,Purrer:2014fza,Purrer:2015tud,Abbott:2016izl,Babak:2016tgq,Bohe:2016gbl}, the latter
based on the effective-one-body~(EOB) formalism~\cite{Buonanno:1998gg,Bernuzzi:2014owa,Bernuzzi:2015rla,
  Hinderer:2016eia}. The weak-field (low-frequency) regime is well described as a combination of point-particle PN
dynamics~\cite{Blanchet:2006zz} with extra finite-size tidal corrections encoded in the tidal Love
numbers~(TLNs)~\cite{PoissonWill}. NR-calibrated phenomenological models are constrained to recover the low-frequency
solutions while correcting the deviations of the higher-order coefficients, which become important in the high-frequency
regime. Thus, any new term included in the PN equations would also propagate to full waveform approximants, possibly in
a contrived and nonlinear way.
 
Tidal deformability terms in the GW phase appear at 5PN order, but are magnified by the fifth power of $GR/(c^2M)$,
where $M$ and $R$ are the stellar mass and radius, respectively. Thus, less compact stars (which are also  more deformable) have a
larger impact in the waveform relative to the point-particle phase.

For nonspinning objects, the TLNs are naturally separated into two classes: those related with induced mass multipole
moments (the so-called \textit{electric} TLNs), and those related to the induced current multipole moments (the
so-called \textit{magnetic} TLNs). When the object is spinning, angular momentum gives rise to spin-tidal coupling and
to a new class of \emph{rotational} TLNs~(RTLNs)~\cite{Pani:2015nua,Landry:2015cva,Landry:2015zfa,Landry:2017piv,
  Gagnon-Bischoff:2017tnz}.

Up to now, TaylorF2 approximants~\cite{Damour:2000zb,Damour:2002kr,Arun:2004hn}, which are based on the PN expansion of
the orbital equations, properly account only for the contribution of the electric TLNs at the leading (5PN) and
next-to-leading~(6PN) order.
On the other hand, tidal terms in EOB models have been partially included up to 
 \comment{7.5PN order, where the tidal-tail terms up to 7.5PN order emerge
naturally from the expansion of the full EOB dynamics~\cite{Damour:2012yf}. EOB models 
including tidal effects have been recently improved by the resummed time-domain 
version~\cite{Nagar:2018zoe}}, but neglecting the effect
of the spin, the magnetic TLNs, and the electric TLNs higher than the quadrupole~\cite{Damour:2012yf}. 

The scope of this work is to quantify the effects of the higher-order tidal terms, namely the magnetic
TLNs~\cite{Binnington:2009bb,Yagi:2013sva,Banihashemi:2018xfb} (whose leading contribution enters at 6PN order), of  the
tidal-tail terms~\cite{Damour:2012yf} (whose leading-order contribution enters at 6.5PN order), and of the recently
computed spin-tidal terms~\cite{Abdelsalhin:2018reg,Landry:2018bil} (which enter at 6.5PN order and are linear in the
binary-component spins).  Although these effects are presumably small, they might be important for several reasons:
(i)~Neglecting them might introduce systematics in the parameter estimation. This is especially important for the
estimate of the tidal deformability, whose relative \snr accumulates mostly at high frequency, where the PN expansion is
poorly convergent. (ii)~Higher-order tidal terms could be used together with simulations to recalibrate effective
models, thus obtaining a more precise approximation of the GW signal at high frequency. (iii)~Spin-tidal couplings may
break some of the degeneracy between tidal and spin effects, thus improving the parameter estimation of both quantities.

\section{Tidal deformations of neutron stars in coalescing binary systems}
The theory of tidal deformation of compact bodies in general relativity has been developed
in~\cite{Flanagan:2007ix,Hinderer:2007mb,Binnington:2009bb,Damour:2009vw}, for nonrotating bodies, and then extended to
rotating bodies in~\cite{Poisson:2014gka,Pani:2015hfa,Pani:2015nua,Landry:2015zfa,Poisson:2016wtv}. This theory has then
been applied to compact binary systems, in order to compute the contribution of the tidal deformation to the emitted
gravitational waveform, in~\cite{Vines:2010ca,Vines:2011ud,Damour:2012yf} for nonrotating NSs, and
in~\cite{Abdelsalhin:2018reg,Landry:2018bil} for rotating stars.

\subsection{Tidal Love numbers of a spinning neutron star}
\label{sec:tln}
When a static, isolated and spherically symmetric object is perturbed by an external tidal field, its mass and current
multipole moments~\cite{Geroch:1970cd,Hansen:1974zz,Thorne:1980ru} (see also~\cite{Cardoso:2016ryw}) are deformed. In
the {\it adiabatic approximation}, in which the external tidal field is (adiabatically) static over the timescale of the
star's response, the mass and current multipole moments (within first-order perturbation theory) are proportional to the
electric and magnetic components of the tidal field. If the NS is not rotating~\footnote{We use the symmetric-trace-free
  (STF) notation~\cite{Thorne:1980ru} where capital letters in the middle of the alphabet $L$, $K$, etc. are shorthand
  for multi-indices $a_1\dots a_l$, $b_1\dots b_k$, etc., and round $(~~)$, square $[~~]$, and angular $\langle~~\rangle$
  brackets in the indices indicate symmetrization, antisymmetrization and trace-free symmetrization, respectively.}
\begin{equation}
Q^L=\lambda_lG^L~~;~~~S^L=\frac{\sigma_l}{c^2}H^L\label{adiab_nonrot}
\end{equation}
where $Q^L$ and $S^L$ are mass and current multipole moments of order $l$, respectively; $G^L$ and $H^L$ are electric and
magnetic tidal tensors of order $l$ evaluated on the star's location; and $\lambda_l$ and $\sigma_l$ are electric and
magnetic TLNs (we follow the notations and conventions of~\cite{Abdelsalhin:2018reg}). We remark that in the nonspinning
case an $l$-pole tidal field can only induce an $l$-pole moment with the same parity. If the object rotates, instead,
moments with different orders $l$ and $l'=l\pm1$ and with opposite parity are coupled to linear order in the spin. For
$l=2,3$, Eqs.~\eqref{adiab_nonrot} generalize to
\begin{align}
Q^{ab} = & \lambda_{2} G^{ab} + \frac{\lambda_{23}}{c^2} J^c H^{abc} \nonumber\\
Q^{abc} = & \lambda_{3} G^{abc} + \frac{\lambda_{32}}{c^2} J^{\langle c} H^{ab \rangle} \nonumber\\
S^{ab} = & \frac{\sigma_{2}}{c^2} H^{ab}  + \sigma_{23} J^c G^{abc} \nonumber \\
S^{abc} = & \frac{\sigma_{3}}{c^2} H^{abc} + \sigma_{32} J^{\langle c} G^{ab \rangle}
\label{eq:adiabatic}
\end{align} 
where we have neglected the multipole moments and the tidal tensors with $l>3$.  In \autoref{eq:adiabatic}, $J^a$ is the
spin vector of the star, and $\lambda_{ll'}$ and $\sigma_{ll'}$ are the
RTLNs~\cite{Pani:2015nua,Landry:2015cva,Landry:2015zfa}. For an $N$-body system, the TNLs and RTLNs are denoted
$\lambda^{(A)}_l$, $\sigma^{(A)}_l$, $\lambda^{(A)}_{ll'}$, $\sigma^{(A)}_{ll'}$, where $A=1,2,\dots$ refers to the
$Ath$ body of the system.

The TLNs (and the RTLNs), computed by employing the relativistic perturbation theory of compact stars, depend on the NS
EoS~\cite{Hinderer:2007mb,Binnington:2009bb,Damour:2009vw,Pani:2015nua,Landry:2015cva,Landry:2015zfa,Gagnon-Bischoff:2017tnz};
indeed, a stiffer EoS corresponds to a more deformable NS, and thus to higher TLNs, whereas a softer EoS corresponds to
lower TLNs and RTLNs.
\subsection{Gravitational waveform of tidally deformed compact binaries up to $\mathbf{6.5}$PN order}
\label{sec:pndesc}
The largest tidal deformation of NSs occurs in the last stages of a compact binary coalescence. In this process, the
binary system emits a strong GW signal, which depends on the TLNs and on the RTLNs of the NSs that are coalescing.

The signal emitted during the inspiral can be described by the PN formalism, recalibrated by fitting unknown (and
possibly resummed) higher-order coefficients to NR solutions. The leading-order contributions of tidal deformation to
the waveform appear at $5$PN order; they do not depend on the NS spin and have been computed
in~\cite{Flanagan:2007ix}. The next-to-leading order contributions, which also do not depend on the spin and appear at
$6$PN order, have been computed in~\cite{Vines:2010ca,Vines:2011ud,Yagi:2013sva}. The (nonspinning) tail component
(which appears at $6.5$PN order) has been computed in~\cite{Damour:2012yf}. Finally, the complete $6.5$PN tidal
waveform, which depends linearly on the NS spin, has recently been computed in~\cite{Abdelsalhin:2018reg}. For the sake
of clarity, we show here the explicit expression for the tidal contribution to the waveform, up to $6.5$PN order,
 \begin{align}
 \psi_T(x) &= \frac{3}{128 \nu x^{5/2}} \Bigg\{ -\frac{39}{2}\tilde\Lambda x^5 \nonumber\\
 & + (\delta\Lambda+\tilde\Sigma) x^6 + (\hat\Lambda+\hat\Sigma+ 
\hat\Gamma+ \hat K ) x^{6.5} +{\cal O}(x^7)\Bigg\}\,, \label{PHASE}
\end{align}
where~\footnote{Note that we slightly changed the notation with respect to~\cite{Abdelsalhin:2018reg}, to be consistent
  with the notation of the LIGO-Virgo papers.}
\begin{align}
 \tilde\Lambda &=\frac{16}{13}\left(\frac{12}{\eta_1}-11\right) \eta_1^5\Lambda_1+
  (1\leftrightarrow 2)\,,\label{defLambda}\\
 \delta\Lambda&=\left( \frac{5095}{28}- \frac{15895}{28 \eta_1} + \frac{5715 
\eta_1}{14} -
  \frac{325 \eta_1^2}{7}  \right)\eta_1^5\Lambda_1 \nonumber\\
  & +(1\leftrightarrow 2) \,,\\
 \tilde\Sigma &= 
  \left( \frac{6920}{7} - \frac{20740}{21 \eta_1} \right) 
  \eta_1^5\Sigma_1 
  +(1\leftrightarrow 2)\,,
\end{align}
\begin{align}
\hat\Lambda &=
  \left[ \left( \frac{593}{4} - \frac{1105}{8 \eta_1} +\frac{567 \eta_1}{8} 
-81 
\eta_1^2 \right) \chi_2\right. \nonumber\\
  &+\left.\left( -\frac{6607}{8} +\frac{6639 \eta_1}{8} -81 \eta_1^2 \right) 
\chi_1 \right] \eta_1^5\Lambda_1+(1\leftrightarrow 2)\,, \label{hatLambdaE}\\
 \hat\Sigma &=\left[\left(-\frac{9865}{3} + \frac{4933}{3 \eta_1} + 1644 
\eta_1 \right) \chi_2 -\chi_1 \right]
 \eta_1^5\Sigma_1\nonumber \\
 &+(1\leftrightarrow 2)\,, \label{hatLambdaM} \\
 \hat K &=\frac{39}{2}\pi \tilde\Lambda\,, \label{hatK} \\
\hat\Gamma &=\frac{\chi_1}{M^4}\left[ \left(  856 \eta_1
 - 816 \eta_1^2 \right){\lambda_{23}^{(1)}} -
 \left(\frac{833 \eta_1}{3} - 278 \eta_1^2 \right) {\sigma_{23}^{(1)}}\right. \nonumber\\
& \left. - \nu \left(272 {\lambda_{32}^{(1)}} -204 {\sigma_{32}^{(1)} }\right)\right]+(1\leftrightarrow 2) \,,
 \end{align}
where $\eta_A = M_A/M$, $M = M_1 + M_2 $ is the total mass of the binary, $M_A$ is the (Newtonian) mass of the $Ath$
body, $\nu = \eta_1 \eta_2$ is the symmetric mass ratio, $\Lambda_A=\lambda_2^{(A)}/M_A^5$,
$\Sigma_A=\sigma_2^{(A)}/M_A^5$ ($A=1,2$), $x=\frac{1}{c^2}(M\omega)^{2/3}$, $\omega$ is the orbital angular velocity,
and $\chi_A = c J_A/M_A^2$ is the dimensionless spin parameter of the $Ath$ object with angular momentum $J_A$ (in
absolute value).
For simplicity, we have used $G=c=1$ units in the above equations; the form of the latter in physical units is given in
Ref.~\cite{Abdelsalhin:2018reg}.
 
As discussed in Ref.~\cite{Abdelsalhin:2018reg}, there seems to exist a conceptual issue related to the inclusion of the
RTLNs in the Lagrangian formulation. Since this problem is still unresolved, in the rest of the paper we shall neglect
the RTLNs, setting $\tilde \Gamma=0$ in the GW phase.

Finally, we added~\autoref{PHASE} to the standard PN point-particle phase~\cite{Blanchet:2006zz} up to $3.5$PN order and
up to linear order in the spin. We neglect quadratic and higher-spin corrections because they are expected to be small
for NS binaries. With this choice, the point-particle phase does not depend on the spin-induced quadrupole moments of
the binary components, which are quadratic in the spin and depend on the EoS.

\section{Statistical analysis}\label{sec:stat}

A figure of merit of GW data analysis is the matched-filter SNR, $\rho$, defined through
\begin{equation}
\label{eq:msnr}
\rho^2=\left( d | h_T\right) = \left( h | h_T	\right) + \left( n | h_T 
\right) \,,
\end{equation}
where $d=h+n$ is a data stream containing a time-domain GW signal $h(\vec{\gamma}_0,\vec{\theta}_0)$, $n$ is a given
realization of the noise, and $h_T(\vec{\gamma}_T,\vec{\theta}_T)$ is a waveform from a given template bank. Each
waveform depends upon a set of $D$-dimensional intrinsic (physical) parameters $\{\vec{\gamma}_0,\vec{\gamma}_T\}$, and
upon a set of extrinsic parameters $\{\vec{\theta}_0,\vec{\theta}_T\}$ that account for angular positions, wave
polarization and distance to the source.
The operator $\left( h | h_T \right)$ defines the overlap between two waveforms,
\begin{equation}
\label{eq:inner}
\left( h | h_T \right)= 4 
\mathcal{R}\int_{f_{min}}^{f_{max}}\frac{\tilde{h}(f)\tilde{h}^*_T(f)}{S_n(f)}
df 
\,,
\end{equation}
with $f_{min}$ and $f_{max}$ the lower and upper cutoff frequencies of the given detector, and $\tilde{h}$ and $S_n(f)$
the frequency domain representation of the signal $h$ and the noise sensitivity curve, respectively.

The extrinsic parameters are irrelevant for waveform modeling purposes since they can naturally  be factored out. Then,
for most of the waveform model computations, \autoref{eq:inner} is usually replaced by the normalized noise-weighted
inner product or match, defined as
\begin{equation}
\label{eq:match}
\mathcal{M}(h(\vec{\gamma}_0),h_T(\vec{\gamma}_T))=\underset{\vec\theta_T}{max}\frac{\left( h | h_T 
\right)}{\sqrt{\left( h | h \right) \left( h_T | h_T 
\right)}} \,,
\end{equation}
where the dependence on the extrinsic parameters is removed by (i) maximizing \autoref{eq:inner} over them and (ii)
normalizing to remove the amplitude scaling. Equation~\eqref{eq:match} provides a useful tool to measure the metric
distance between two waveform representations, since $\mathcal{M}\in[0,1]$, with $\mathcal{M}=1,0$ being a perfect and a
zero match, respectively. In general, $\mathcal{M}$ is used as an indicator of the performance of waveform models and,
for high-SNR and Gaussian noise, it may be used to provide an estimate of the systematic errors produced by the
different waveform representations.

On the other hand, the parameter estimation of GW signals is based on the application of Bayesian information theory to the
observed data streams. To do so, we have to compute the posterior distributions in which the data streams are matched to
the waveform template banks \cite{veitch:2014wba},
\begin{equation}
\label{eq:probBay}
p(\vec{\gamma}_T|d )\propto p_0(\vec{\gamma}_T) \mathrm{e}^{-\frac{\left( d- h_T 
|d-h_T \right)}{2}}\,,
\end{equation}
where $p_0$ is the prior distribution of the intrinsic parameters $\vec\gamma_T$. For high SNR, Gaussian noise and
assuming flat priors, \autoref{eq:probBay} may be substantially simplified by neglecting the noise-related factors. In
this case, one can express the multivariate posterior distribution around the true parameters $\vec{\gamma}_0$ as (see,
e.g., Appendix~G of Ref.~\cite{Chatziioannou:2017tdw})
\begin{equation}
\label{eq:probFMmr}
p(\vec{\gamma}_T)\propto \exp\left(-\rho^2(1-\mathcal{M}(h|h_T))\right)\,,
\end{equation}
where we have removed the arguments to simplify the notation. The above equation allows us to describe completely the
statistics in terms of the \snr and the match $\mathcal{M}(h,h_T)$. In other words, for a given \snr $\rho$ and a given
template $h_T$, the mismatch $1-\mathcal{M}$ determines the probability distribution around the true values
$\vec{\gamma}_0$. Note that the true parameters given by $\vec{\gamma}_0$ do not correspond to the recovered ones
$\vec{\gamma}_T$ unless the real (injected) waveform and the template bank used are equal, $h=h_T$. This may insert
non-negligible systematic errors that in some cases may compete in significance with the statistical ones. Thus, if we
replace $h$ by a given waveform template, \autoref{eq:probFMmr} allows us to estimate the impact of using one or another
waveform template in our parameter estimation. We evaluate these effects in~\autoref{sec:res}, including the PN
corrections described by~\cite{Abdelsalhin:2018reg}.

An alternative approach is based on the Fisher-information matrix (FIM)
approximation~\cite{Cutler:1994ys,Vallisneri:2007ev,Chatziioannou:2017tdw}, which is known to be valid for large values
of the SNR and when the noise is mostly Gaussian. In this case, it turns out that the probability of having each of the
reconstructed parameters shifted by $\Delta \gamma_i=(\vec\gamma_T-\vec{\gamma}_0)_i$ from the real values is given by
\begin{equation}
\label{eq:probFMs}
p(\vec{\Delta \gamma})\propto \exp\left({-\frac{1}{2}\Gamma_{ij} \Delta \gamma_i \Delta \gamma_j}\right) \,,
\end{equation}
where
$ \Gamma_{ij}=\left(\frac{\partial h_{T}}{\partial\gamma_{T\,i}},
 \frac{\partial h_{T}}{\partial\gamma_{T\,j}}\right)\bigg{|}_{\vec{\gamma}_T=\vec{\gamma}_0}$
 is the FIM. Then, we can compute the value of the $D$-dimensional posterior when each of the reconstructed parameters
 is $n\,\sigma$ away from the maximum-likelihood ones as
\begin{equation}
\label{eq:probFMs2}
p(\vec{\gamma}_{n\,\sigma})\propto \exp\left({-\frac{1}{2} \Gamma_{ij} n^2 \sigma_{\gamma_i} \sigma_{\gamma_j} }\right) \,,
\end{equation}
where $\sigma_{\gamma_i} = \sqrt{\Gamma^{-1}_{ii}}$ is the statistical error on the $ith$ parameter $\gamma_i$ \footnote{A weaker requirement would be looking for the \emph{global} $n\, \sigma$ confidence level hypersurface. Since in the FIM approximation the posterior distribution is Gaussian, this surface is a $D$-dimensional ellipsoid, and
\begin{equation}
p(\vec{\gamma}_{n\,\sigma})\propto \exp\left({-\frac{1}{2} r^2 }\right) \,,
\end{equation}
where $r$ is the Mahalanobis distance
\begin{equation}
r^2 = \phi(c(n),D) \,,
\end{equation}
with $\phi$ the inverse of the cumulative distribution of the $\chi^2$-distribution with $D$ degrees of freedom and
$c(n)$ the probability of falling inside the $n\, \sigma$ confidence region ($c(1) \sim 0.68, \, c(2) \sim 0.95, \,
\text{etc}.$).}. Neglecting the correlation among the parameters,~\footnote{This assumption is justified by the fact
  that we are interested only in the weighted-tidal deformability $\tilde{\Lambda}$ parameter, which at so high SNR is weakly
  correlated to the other parameters (cf. \autoref{sec:res}).} \autoref{eq:probFMs2} reduces to
\begin{equation}
\label{eq:probFMs3}
p(\vec{\gamma}_{n\, \sigma})\propto \mathrm{e}^{-\frac{D}{2}n^2} \,.
\end{equation}
Thus, by equating \autoref{eq:probFMmr} and \autoref{eq:probFMs3} one gets
\begin{equation}
\label{eq:ind}
(h - h_T | h - h_T ) \approx   2 \rho^2 (1-\mathcal{M}) 
\approx D \,n^2  \,.
\end{equation}
The above expression allows defining the well-known \textit{distinguishability criterion} between two waveform
models~\cite{Lindblom:2008cm}. In other words, it allows us to estimate the minimum \snr required to distinguish two
waveform models within a certain ${n=\Delta \gamma}/{\sigma}$ significance, with the latter ratio equal to unity to
distinguish two models with $1\sigma$ significance, for instance. We use this definition in \autoref{sec:res} as a
quantitative indicator of the impact of the new terms described in the Introduction where ${D=6: \left\lbrace M, \nu,
  \chi_1,\chi_2, \Lambda_1, \Lambda_2 \right \rbrace }$, where we set ${n = 1.64}$ to get the results at the $90\%$
credible level.
\section{Impact of the higher-order tidal terms in 
the GW phase}\label{sec:qual}
The new terms considered in Sec.~\ref{sec:pndesc} modify the waveform at high PN order. This implies that their effects
gain importance as the signal approaches the high-frequency regime, possibly probing a region where current
gravitational detectors are less sensitive. In general the impact of these terms will depend on the source parameters
and on the merger frequency relative to the detector sensitivity. Fortunately, the parameter range of NS mergers appears
to be much reduced with respect to the binary black hole case which simplifies the task of exploring the full BNS parameter space. The astrophysically relevant BNS systems are expected to have a total mass $M$ that
lies in the range $[2.5,4]M_\odot$, the mass ratio is expected to be $ M_1/M_2\in[1,2]$, individual spins are expected
to be small\footnote{Although the distribution of NS spins is uncertain, old NSs in the late stages of a binary inspiral
  are expected to rotate rather slowly. The  fastest spinning NS observed so far in a compact system is the most
  massive component of the double pulsar system PSR~ J0737-3039A~\cite{2003Natur.426..531B}, with a spin period of
  $\approx 23\,{\rm ms}$, which corresponds to $\chi\sim0.02-0.05$, depending on the
  EoS~\cite{PhysRevD.86.084017,2013PhRvD..88b1501K} ($\chi\sim0.02$ for APR~\cite{Akmal:1998cf} EoS). Such a rotation rate
  is not expected to decrease substantially as this system approaches the merger (see Ref.~\cite{Dietrich:2015pxa} for a
  discussion). On the other hand, the observation of numerous isolated millisecond pulsars suggests that spin rates as
  high as $\chi\sim0.1$~\cite{Dietrich:2015pxa} might be found also in BNS systems.}, $\chi_{1,2} \lesssim 0.05$, while
the recent LIGO-Virgo constraints on the deformability parameters are $\tilde\Lambda<800$ at the $90\%$ credible
level~\cite{TheLIGOScientific:2017qsa,De:2018uhw,Abbott:2018exr,Abbott:2018wiz}. Then, before running any expensive
parameter-estimation analysis, we provide qualitative intuition on the importance of the terms considered in
Sec.~\ref{sec:pndesc} by showing how they affect the GW phase for different masses, mass ratios, spins, and tidal
deformability coefficients.

\subsection{Relevance of the magnetic TLNs} 
Let us start by discussing the magnitude of the magnetic TLN term, $\tilde\Sigma$, entering~\autoref{PHASE} at 6PN
order. This term arises from the odd-parity sector of the perturbation equations \cite{Binnington:2009bb,Damour:2009vw}
thus being in principle independent from the electric TLN term. However, there exist some approximate EoS-independent
relations, $\Sigma_i=\Sigma_i(\Lambda_i)$, that connect the magnetic TLNs to the standard electric
TLNs~\cite{Yagi:2013sva}. Notice that the magnetic TLNs depend on the properties of the
fluid~\cite{Landry:2015cva,Landry:2015snx,Gagnon-Bischoff:2017tnz}. In particular, the magnetic TLNs for irrotational
fluids or for static fluids have the opposite sign and the quasiuniversal relation also depends on the fluid
properties.

We have recently revised the properties of odd-parity perturbations and of the magnetic TLNs of a NS~\cite{Pani:2018inf}. Our analysis confirms the discussion in Ref.~\cite{Landry:2015cva} by showing
that assuming an irrotational fluid provides a more realistic description of the fluid dynamics, since dynamical
odd-parity perturbations enforce irrotationality in the stationary limit~\cite{Pani:2018inf}. Nevertheless, for the
sake of completeness, here we shall consider both cases and we use the following fitting formula 
\begin{equation}
\label{fit}
\log (\pm\Sigma_i)= \sum_{n=0}^5 a_n \left(\log 
\Lambda_i\right)^n\,,
\end{equation}
where $i=1,2$ for the two bodies, and the plus or minus sign is for static or irrotational fluid, respectively. The
form of the above fit is the same as that of Ref.~\cite{Yagi:2013sva}, whereas the coefficients have been computed in
Ref.~\cite{Pani:2018inf} and are given in Table~\ref{tab:magnetic} for the cases of irrotational and static fluids,
respectively. Our results agree with those of Ref.~\cite{Gagnon-Bischoff:2017tnz} and with the revised ones in
Ref.~\cite{Yagi:2013sva} in the relevant regimes~\cite{Pani:2018inf}. We also checked that the analysis presented below
is insensitive to the small differences between different fitting functions.
\begin{widetext}
\center
\begin{table}
\label{tab:magnetic}
\begin{tabular}{|c|cccccc|}
\hline
fluid		 & $a_0$ & $a_1$ & $a_2$ & $a_3$ & $a_4$ & $a_5$\\
\hline
irrotational 	 & $-2.03$ & $4.87 \times 10^{-1}$ & $9.69 \times 10^{-3}$ & $1.03 \times 10^{-3}$ & 
$-9.37 \times 10^{-5}$ & $2.24 \times 10^{-6}$\\
static		 & $-2.66$ & $7.86 \times 10^{-1}$ & $-1.00  \times 10^{-2}$ & $1.28 \times 10^{-3}$ & 
$-6.37 \times 10^{-5}$ & $1.18 \times 10^{-6}$ \\
\hline
\end{tabular}
\caption{Coefficients of the fitting formula~\eqref{fit} describing the approximated EoS-independent relation between
  magnetic and electric TLNs. We consider both irrotational and static fluids~\cite{Pani:2018inf}.}
\end{table}
\end{widetext}

Equation~\eqref{fit} provides a mapping from $\Lambda_{1,2}$ to $\Sigma_{1,2}$ which is accurate at the level of a few
percent~\cite{Yagi:2013sva,Gagnon-Bischoff:2017tnz,Pani:2018inf}. Using this relation allows one to remove the explicit
dependence on the magnetic TLNs in the waveform.  In this way the final phase~\eqref{PHASE} depends only on the electric
TLNs $\Lambda_{1,2}$, on the masses $M_{1,2}$, and on the spins $\chi_{1,2}$.
\begin{figure}[!htb]
\includegraphics[width=\columnwidth]{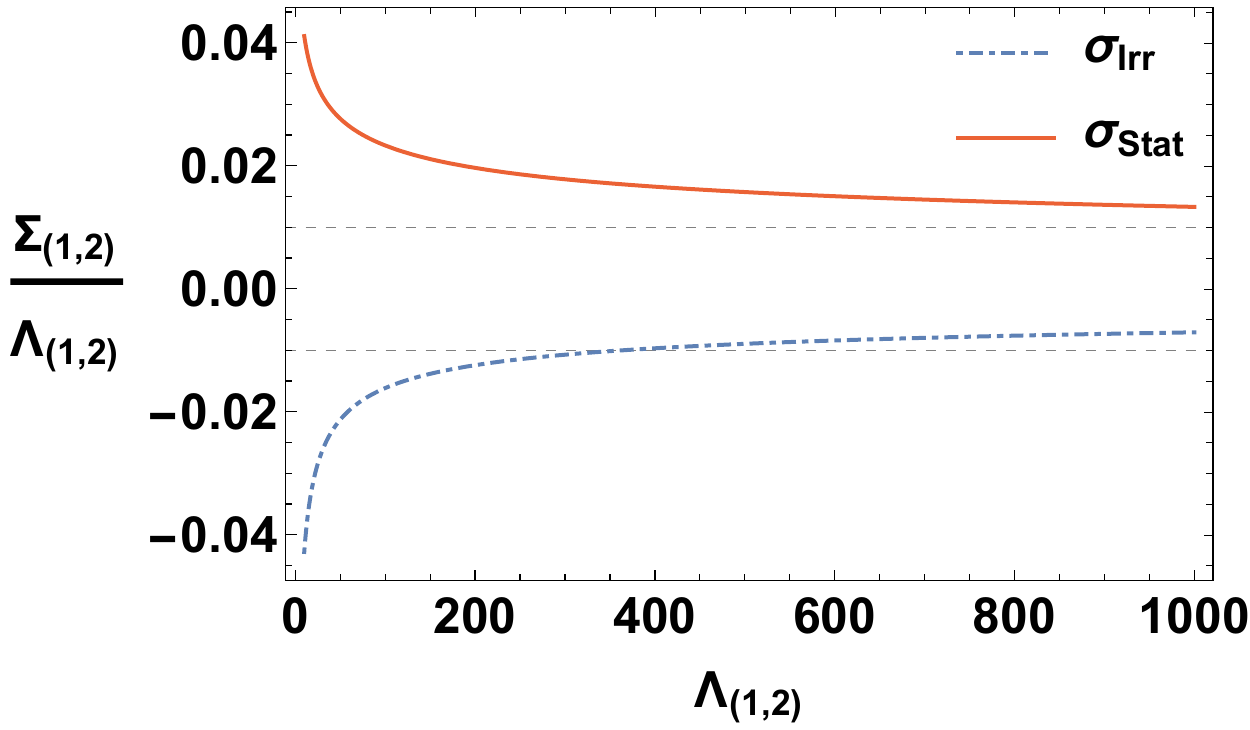}
 \caption{Approximated universal relations between the electric ($\Lambda_i$) and the magnetic ($\Sigma_i$) quadrupolar
   TLNs as proposed in~\cite{Yagi:2013sva}. The dashed lines define the points in the parameter space where the magnetic
   TLN contribution reaches $1\%$ with respect to standard electric TLNs. The magnetic TLNs depend on the properties
   of the fluid~\cite{Landry:2015cva,Landry:2015snx,Gagnon-Bischoff:2017tnz}; we show the cases of irrotational and of
   static fluids.
 \label{fig:magnetic_Vs_electric}
 }
 \end{figure}

In \autoref{fig:magnetic_Vs_electric} we show the approximate EoS-independent relations between the (quadrupolar)
electric and magnetic TLNs for irrotational and static fluids. For typical values of the compactness of a NS, the ratio
$\Lambda_i/\Sigma_i\approx \pm 100$, where the plus and minus signs refer to static and irrotational fluids,
respectively. This anticipates that the 6PN order coefficient in the tidal phase is dominated by the next-to-leading
corrections proportional to $\Lambda_{1,2}$, rather than by the terms proportional to $\Sigma_{1,2}$ that enter at the
same PN order. As shown in \autoref{fig:magnetic_Vs_electric}, the ratio $\Sigma_i/\Lambda_i$ depends only mildly on
$\Lambda_i$ and, when $\Lambda_i\gtrsim200$, the relation $\Sigma_i(\Lambda_i)$ is approximately linear.
As already mentioned, assuming static or irrotational fluid yields opposite magnetic corrections to the tidal
phase. Static fluids yield magnetic TLNs that appear with the same global sign as the $\Lambda_i$ and the $\delta
\Lambda$ terms (i.e.,  attractive tidal effects in the two-body dynamics). This would (slightly) increase the overall
impact of the tidal effects at the next-to-leading order. On the other hand, the more realistic case of irrotational fluid yields magnetic TLNs that act in the opposite direction (repulsive tidal effects), thus inducing a partial
screening of the subleading tidal terms. This qualitative analysis suggests that neglecting the magnetic contribution
may lead to a (small) bias in the estimate of $\Lambda_i$ that will be of opposite sign depending of the type of fluid
considered. We explore the relevance of these effects in~\autoref{sec:res}.

\subsection{Relevance of the spin-tidal couplings} 

\begin{figure}[!htb]
 \includegraphics[width=\columnwidth]{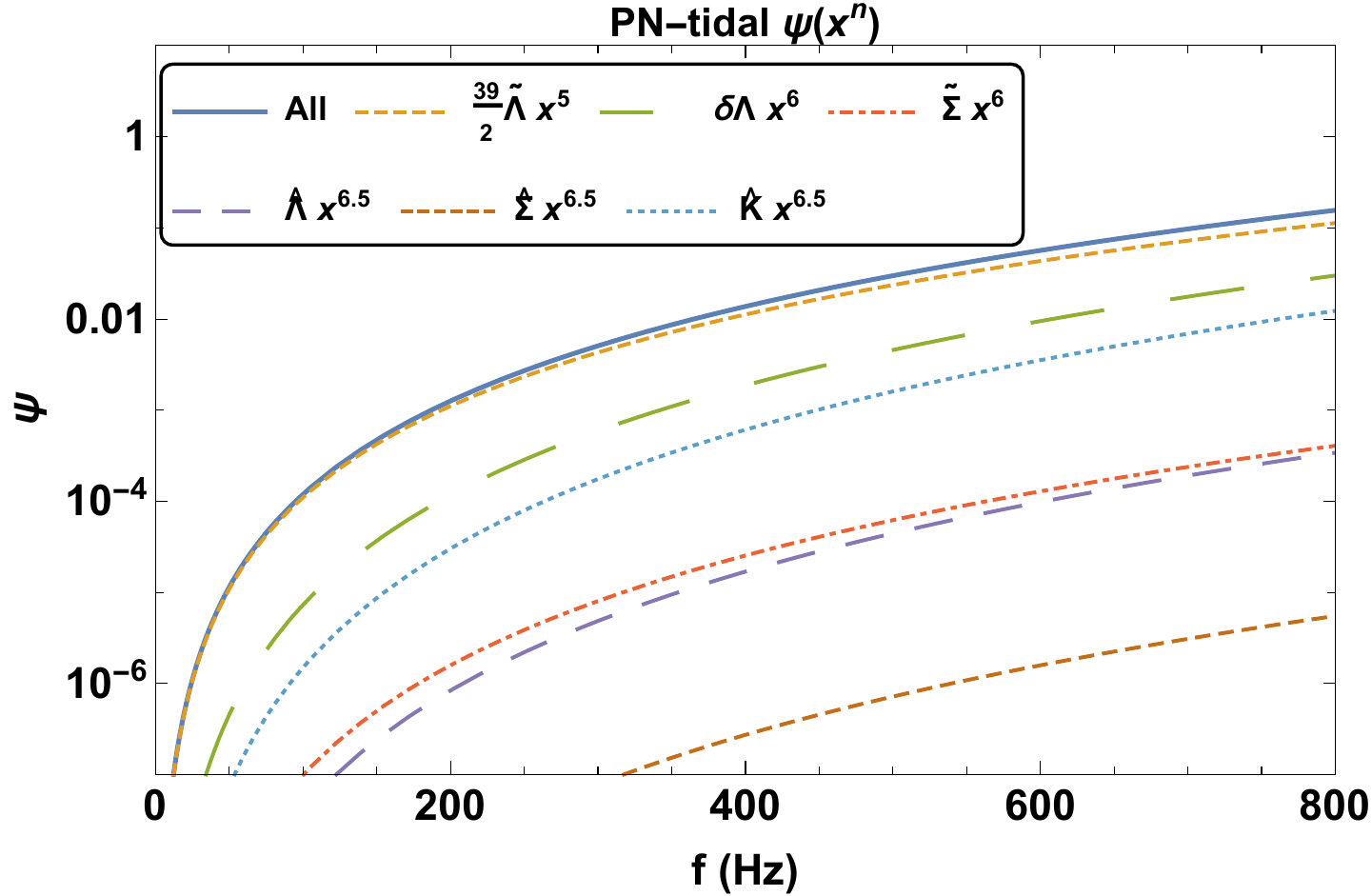}
 \caption{Each of the tidal contributions from \autoref{PHASE} to the GW phase (in absolute value) as a function of the
   GW frequency. We considered an equal-mass binary with total mass $2.8M_{\odot}$, with $\Lambda_1 = \Lambda_2 = 300$,
   and spins $\chi_1=\chi_2=0.05$. The contribution from the spin-tidal couplings at $6.5$PN order $\hat \Lambda$ scales
   linearly with the spin. } \label{fig:6PNall}
  \vspace*{-\baselineskip}
 \end{figure}
 
In \autoref{fig:6PNall} we present the individual contributions to the GW phase of each of the tidal terms described in
\autoref{PHASE} along the relevant LIGO-Virgo frequency domain. To do so, we consider a system with physical parameters
roughly compatible with GW170817: an equal-mass binary with total mass $2.8M_{\odot}$, electric TLNs $\Lambda_1 =
\Lambda_2 = 300$, and spin parameters $\chi_1=\chi_2=0.05$, in order to show the effects of the 6.5PN order spin-tidal
coupling terms $\hat \Lambda$. As expected, the leading-order 5PN order term $\tilde \Lambda$ dominates the GW tidal
phase by more than an order of magnitude with respect to higher-order terms. The next term in order of importance is the
6PN order-electric term $\delta \Lambda$, which contributes on average around  $20 \%$ of the total tidal phase
evolution. Indeed, this was the highest PN tidal term accounted for in the analysis of GW170817~\cite{TheLIGOScientific:2017qsa}.

The next term in order of relevance is the 6.5PN order tidal tail $\hat K$, whereas the spin-tidal term $\hat \Lambda$
is significantly less dominant: its relative contribution is smaller than the total tidal phase by about $2$ orders of
magnitude and also contributes about  $3\%$ with respect to the total 6.5PN order coefficient. This suggests that it might
be safely neglected for binaries with $\chi_i\approx0.05$. On the other hand, this term grows linearly with the spin so
that it might become important if highly spinning NS binaries exist in nature and for high \snr scenarios.
Finally, the lowest contributions come from the magnetic \comment{TLNs~\cite{Damour:2009vw},  which agree with the estimates obtained by~\cite{Bernuzzi:2014owa}}. This has to do with the small ratio between the magnetic
and the electric TLNs shown in \autoref{fig:magnetic_Vs_electric}.

We quantify the above expectations in~\autoref{sec:res}. We note that the higher-order tidal terms have a simple
dependence across the parameter space of the binary. Thus, they can  easily be computed for different values of
$\Lambda_i$, $M_i$, and $\chi_i$.

\begin{figure}
 \includegraphics[width=\columnwidth]{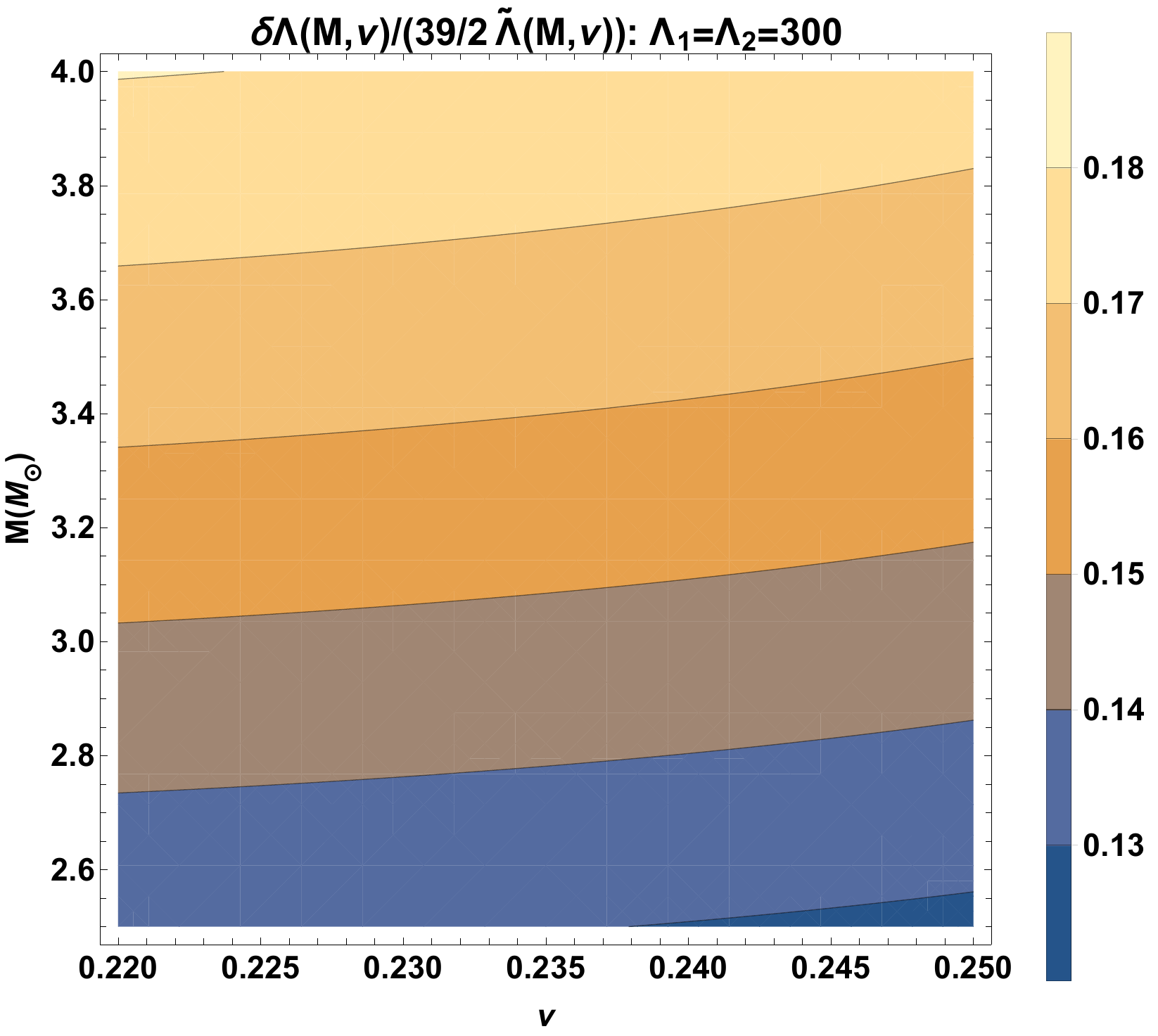}
 \caption{Absolute value of the ratio between the leading-order and the next-to-leading order coefficients,
  ${2/39\,\delta\Lambda/\tilde\Lambda}$, at $f=300\,{\rm Hz}$ as a function of the total mass $M$ and the mass ratio
  $\nu$. We set $\Lambda_1 =\Lambda_2=300$, consistent with
  GW170817~\cite{TheLIGOScientific:2017qsa,De:2018uhw,Abbott:2018exr,Abbott:2018wiz}.
 \label{fig:coefrat65}
\vspace*{-\baselineskip}
 }
\end{figure}
\begin{figure}[!htb]
  \includegraphics[width=\columnwidth]{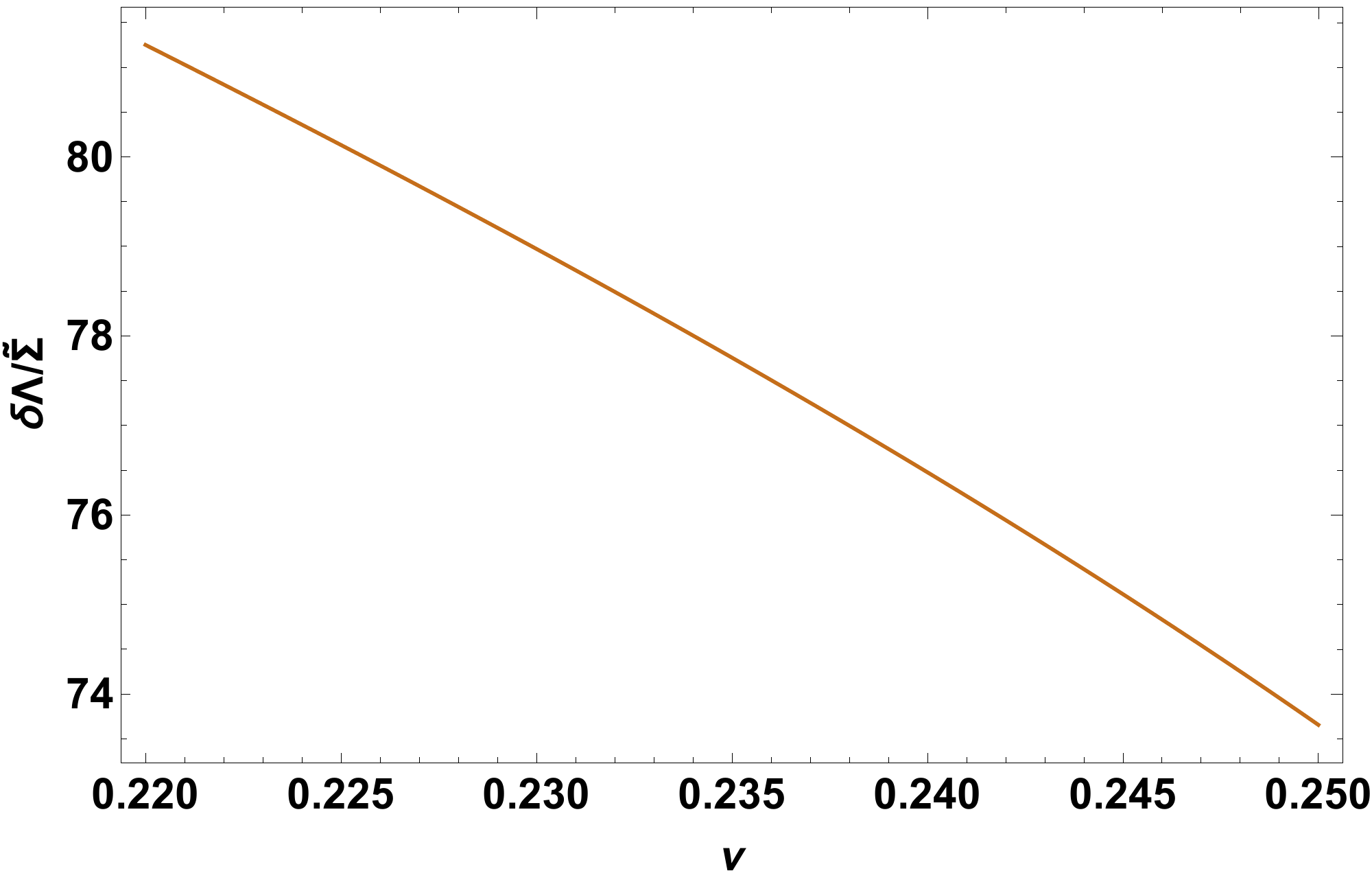}
 \caption{Ratio (in absolute value) between the two next-to-leading order contributions, $\delta\Lambda/\tilde
  \Sigma$, as a function of the symmetric mass ratio $\nu$ for $\Lambda_1 =\Lambda_2=300$ and for irrotational fluids. Notice that the dependence
  on the total mass scales out since we are comparing equal-order coefficients.}
 \label{fig:coefrat66}
\vspace*{-\baselineskip}
\end{figure}
For example, we can explore the ratio of the dominant 5PN-6PN order factors $\tilde \Lambda$ and $\delta \Lambda$ as a
function of the total mass $M$ and the mass ratio $\nu$. We show this in \autoref{fig:coefrat65} for a GW170817-like
event with $\Lambda_1=\Lambda_2=300$. The 6PN order terms $\delta \Lambda$ do not contribute by more than $20\%$
relative to the leading-order term in the whole parameter space.  We observe that the parameter that mostly affects this
ratio is the total mass $M$, being the mass ratio subdominant. As we explain in the next lines, this ratio increases
linearly with $\tilde\Lambda$. This implies that, considering the recent constraint $\tilde\Lambda \lesssim
800$~\cite{TheLIGOScientific:2017qsa,Abbott:2018exr}, this ratio could increase by a factor $\sim 3$, thus allowing for
a maximal contribution of about $40\%$ in ultrahigh mass ($4M_\odot$) NS mergers and of about $30\%$ in ordinary
binaries.

In addition, in \autoref{fig:coefrat66} we show the ratio between the coefficients $\delta\Lambda$ and $\tilde\Sigma$
for a $\Lambda_1=\Lambda_2=300$ binary. The ratio $\delta\Lambda/\tilde\Sigma$ does not depend strongly on the mass
ratio (since we are comparing PN terms at the same order the total mass scales out), thus being dominated by the
different scale between the electric and magnetic TLNs.
\section{Results}\label{sec:res}
To quantify the effect of the 6PN order magnetic TLNs and the spin-tidal 6.5PN order contributions,
we adopt an analysis based on the FIM introduced in~\autoref{sec:stat}. In particular, our analysis is valid
for high-\snr and Gaussian noise. As we shall verify \textit{a posteriori}, the former assumption is anyway necessary,
since the effect of higher-order tidal terms is typically small. 

We take the only BNS event observed so far by the LIGO-Virgo Collaboration, GW170817, as a reference. This event,
observed with a \snr of $\rho = 32.4$, was consistent with a BNS system with masses compatible with
$M_1 \sim M_2 \sim 1.4 M_\odot$, with spins of the components compatible to zero, $\chi_{1,2} \sim 0$, and the $90\%$
credible intervals on the combined electric tidal deformability $\tilde\Lambda$ have been recently constrained to lie
within $\sim [70,800]$ \cite{Abbott:2018wiz,De:2018uhw} with the median value being $\tilde{\Lambda}=300$. \comment{Though the
observation of such a high \snr event was rather unlikely considering the previous event rate
predictions~\cite{Abbott:2016ymx}, the inclination reported by the LIGO-Virgo Collaboration~\cite{TheLIGOScientific:2017qsa} allows for a nonzero inclination angle with respect to the
Earth's observation line (${i \sim 30º}$). This could have been reduced by a factor of approximately $1.15$ the total \snr of
this event relative to the \snr that the same source would have produced if oriented face on (see, e.g., ~\cite{Sathyaprakash:2009xs}).
Based on the above discussion}, we consider two different scenarios: 
the \textit{standard} scenario where the physical parameters are taken to be 
those compatible with GW170817, that is, $\rho=32.4$, $M_1=M_2=1.4M_\odot$, 
$\tilde{\Lambda}=300$; and a more \textit{optimistic} scenario in which we consider 
the hypothetical case of observing an event with the physical parameters compatible with GW170817,  \comment{but in a face-on orientation, i.e., with $\rho=37$,~\footnote{This factor is easily obtained from the $_{-2}Y^{22}(i=30,\phi)$ spherical harmonic.} and fixing} 
the maximum value allowed by the LIGO-Virgo posterior 
distributions. The optimistic scenario maximizes the detectability of the tidal 
effects considered in this work. Finally, we consider two detectors: (i) LIGO in 
its O1 configuration~\cite{Aasi:2013wya}, known as \aLIGOe, with $\left\lbrace 
f_{min},f_{max}\right\rbrace= \left\lbrace 23,2048\right\rbrace\,{\rm Hz}$; and 
(ii) the planned third-generation detector Einstein 
Telescope (ET)~\cite{Hild:2009ns,Sathyaprakash:2011bh,Sathyaprakash:2012jk} in its 
\ET configuration\footnote{The \textit{D} stands for the so-called xylophone 
configuration that is expected to be more sensitive at low frequencies than 
other alternative ET designs as the V-shaped configuration 
ET-B~\cite{Hild:2010id,Sathyaprakash:2011bh}.} with $\left\lbrace 
f_{min},f_{max}\right\rbrace= \left\lbrace 1,2048\right\rbrace\,{\rm Hz}$, for 
which 
current prospects anticipate a sensitivity gain of  factor \comment{$\approx 45$} compared to 
\aLIGOe. The correcting factors to translate a LIGO-Virgo event to an ET one are described in 
Appendix~\ref{sec:appdx}.\footnote{The geometrical factors to transform from a 
LIGO-type observatory to ET are explained in \cite{Hall,Hild:2010id}.}

\subsection{Impact of 6PN order magnetic terms on TaylorF2 approximants}
\label{sub:6pn}
Systematic errors on GW parameter estimation are induced by the incompleteness of the waveform template banks. This may
produce an artificial bias with respect to the true parameters that in some cases may overtake the statistical
uncertainty driven by the dectector's noise. Here we evaluate the impact of neglecting the 6PN order magnetic coefficient
$\tilde\Sigma$ by comparing three different sets of analytic waveforms. As previously discussed, we consider each
magnetic TLN $\Sigma_{i}$ to be related to the electric $\Lambda_{i}$ through the universal relations shown in
Table~\ref{tab:magnetic}. Then, we explore the possibility of observing two simulated GW triggers, tagged as $h_{st}$
and $h_{irr}$, that are match filtered with a waveform template bank $h_0$. The triggers and the template are defined as
\begin{itemize}
\item{$h_0$: TaylorF2 waveform template bank given by \autoref{PHASE}, truncated at 6PN order, and with vanishing
  magnetic TLNs, i.e. $\tilde\Sigma=0$, as in \cite{TheLIGOScientific:2017qsa,De:2018uhw,Abbott:2018wiz};}
\item{$h_{st}$: GW signal consistent with a TaylorF2 model truncated at 6PN order, with magnetic TLNs included assuming
  a static fluid;}
\item{$h_{irr}$: GW signal consistent with a TaylorF2 model truncated at 6PN order, with magnetic TLNs included assuming
  an irrotational fluid.}
\end{itemize}
To produce the posterior distribution~\ref{eq:probFMmr}, we first need to compute the match \ref{eq:match} between our
\textit{magnetic} GW signals $h_{st,irr}$ and our reference waveform template bank $h_0$. We set the masses and spins of
the GW signals $h_{st,irr}$ and the template bank waveforms $h_0$ to ${M_1=M_2=1.4M_\odot}$ and ${\chi_1=\chi_2=0}$,
respectively. In addition, we consider two possible values (injections)
$\tilde\Lambda_0=300,800$ for the tidal deformability of our simulated GW events
$h_{st,irr}$, consistent with our \textit{standard} and \textit{optimistic} 
scenarios. 
We also assume that the two NSs are described by the same EoS so that, since $M_1=M_2$, we have $\Lambda_1=\Lambda_2$
and $\Sigma_1=\Sigma_2$. Then, by varying $\Lambda_1$ such that $\tilde\Lambda\in[0,2000]$, we generate a
$\tilde\Lambda$-dependent distribution for the match $\mathcal{M}(h_0(\tilde\Lambda)|h_{st,irr}(\tilde\Lambda_0))$, that
is translated to $p(\tilde\Lambda)$ by means of \autoref{eq:probFMmr}. We repeat the analysis for \aLIGOe and \ET noise
sensitivity curves.
 \begin{figure}[!htb]
 \includegraphics[width=\columnwidth]{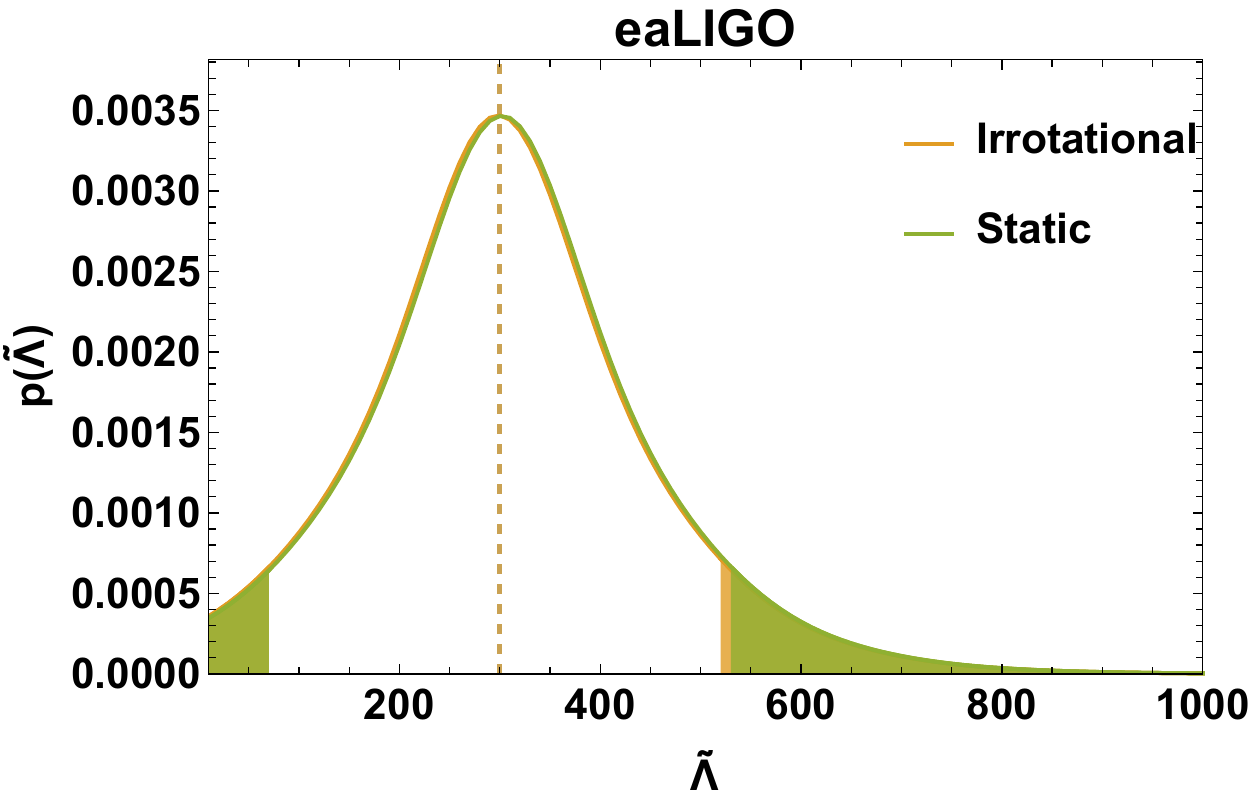}
 \caption{Probability distributions resulting from the match between $h_0$ and $h_{st,irr}$ for an injected value
   $\tilde\Lambda_0=300$ and $\rho=32.4$. The offset produced by neglecting the magnetic TLNs, both static and
   irrotational, is shown to be negligible for events compatible with GW170817 and the \aLIGOe sensitivity curve. The
   solid area defines the region out of the $90\%$ $\tilde\Lambda$ credible intervals.}
 \label{fig:6PN300eL}
 \end{figure}

 \begin{figure}[!htb]
 \includegraphics[width=\columnwidth]
{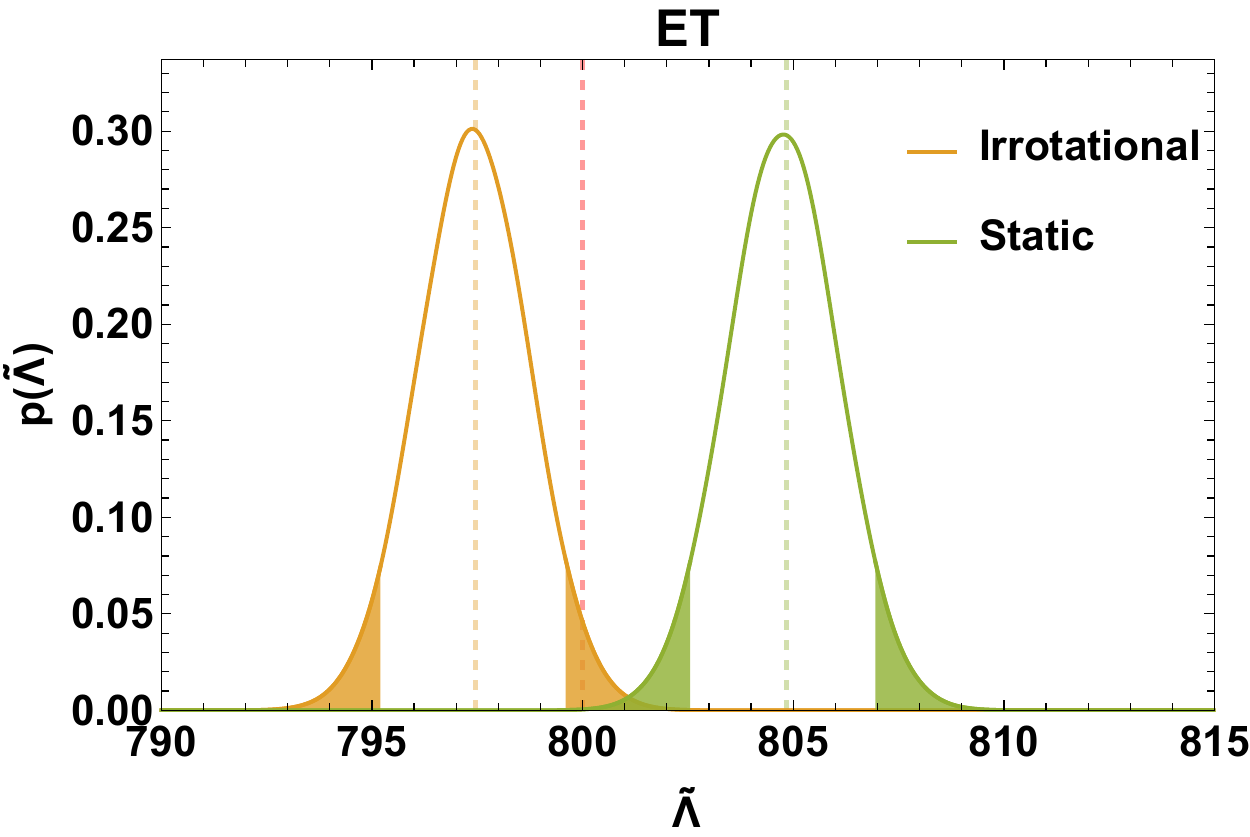}
 \caption{Same as in \autoref{fig:6PN300eL} but for our \textit{optimistic} scenario and assuming a detection with
   \ET. The vertical dashed line in red corresponds to the position of the injected value $\tilde\Lambda$, while the
   orange and the green ones correspond to the peak value for irrotational and static fluids, respectively. The solid
   areas (orange and green) define the region out of the $90\%$ credible intervals.
 \label{fig:6PN800ET}
 }
 \end{figure}
In \autoref{fig:6PN300eL} we show the probability distributions obtained in our \textit{standard} scenario by matching
our two GW events $h_{st,irr}$ to the waveform template bank $h_0$. We observe that the posterior distributions obtained
for $h_{st,irr}$ are compatible with the injected value $\tilde\Lambda_0=300$, thus not revealing any sensitive offset
with respect to the zero-magnetic model $h_0$. The differences on the peaks obtained for static (green) and irrotational
(orange) fluids are on the order of ${|\tilde\Lambda_0-\tilde\Lambda_{st,irr}| \sim 1}$, where the solid area determines
the $90\%$ credible intervals. This is consistent with the results of \autoref{fig:6PNall}, where the $\tilde\Sigma$
contribution appears to be even more subdominant than the $6.5$PN order tidal-tail effect. Moreover, considering that
these effects do not vary significantly in the unequal-mass case (\autoref{fig:coefrat66}) and that their dependence on
the total mass is exactly zero (\autoref{fig:coefrat66}), we do not expect to observe any significant gain for a
different point across the parameter space. Therefore, these results suggest that the effect of magnetic TLNs is
negligible for the measurement of the tidal deformability and that measuring the magnetic TLNs independently with
\aLIGOe will be very unlikely. Unfortunately, this also prevents the constraint of the dynamical properties of the NS fluid
(i.e., static versus irrotational) according to a putative measurement of the sign of $\Sigma_i$.
We note that if these terms are negligible within our simplified analysis (in which the only parameter that is varied is
$\tilde\Lambda$), they would be even more difficult to measure within a rigorous and more expensive multidimensional
Bayesian analysis performed on $h_{st,irr}$.

Finally, we can also carry out the same analysis in our \textit{optimistic} scenario (that is, $\tilde\Lambda=800$) and
assuming the \ET noise sensitivity curve. This gives an SNR larger than in the standard scenario by roughly a factor of  \comment{$55$; i.e., we set $\rho=1750$, which comes from observing 
GW170817 with \ET (see Appendix~\ref{sec:appdx})}. From
\autoref{eq:probFMmr}, in the large-$\rho$ limit the FIM errors scale as $\sigma_i\propto 1/\rho$. This implies that any
gain on the SNR will sharpen our posterior distributions around the best-likelihood values that, in general, will be
different for different waveform approximants. Thus, in the optimistic scenario the offset between the recovered tidal
deformability $\tilde\Lambda$ given by $h_{st,irr}$ and the injected value $\tilde\Lambda_0$ should be larger than in
the standard scenario assuming a GW170817-like detection with LIGO-Virgo.

In \autoref{fig:6PN800ET} we show the posterior distributions generated when assuming static and irrotational
fluids. Notice that the displacement between distributions is larger than in the \textit{standard} scenario, as
well as the $90\%$ credible levels delimited by the orange and green solid areas. The sign of the offset is directly
correlated with the sign of the magnetic TLNs. In the case of irrotational fluids, the magnetic TLNs enter at lower
order than (but with the sign opposite to) the electric TLNs, thus tending to decrease the tidal effects.  This induces
a small underestimation of $\tilde\Lambda$ with respect to the injected value $\tilde\Lambda_0$,  the differences being
larger as one increases the injected value $\tilde\Lambda_0$.  The opposite happens for static fluids: in this case the
sign of electric and magnetic TLNs is the same, leading to an increase of the tidal effects, and thus inducing a small
overestimate of $\tilde\Lambda$. \comment{The offset between the distributions is  sufficient to place the peak values outside of the credible regions.  In other words, the differences may be marginally observable as long as the optimistic scenario is considered. Therefore, based on the above analysis, we can estimate that the error induced by not including tidal-magnetic effects in current waveform
models~\cite{Vines:2011ud,Dietrich:2017aum} (both NR-calibrated and analytical ones) will not be observed for the
next BNS observations with 2G gravitational wave detectors, but they may produce a not 
negligible impact on 3G detectors such as ET.}
\subsection{Impact of 6.5PN order terms on TaylorF2 approximants}
\label{sub:6pn2}

Following the discussion of the previous section, let us now consider the $6.5$PN order terms, i.e., the tidal-tail term
and the tidal-spin coupling. We quantify the magnitude of these terms by performing the same match/distinguishability
analysis previously discussed, but now considering a spinning binary. For simplicity, and because they are more
realistic~\cite{Landry:2015cva,Pani:2018inf}, we only consider the magnetic TLNs arising from an irrotational
fluid.\footnote{At any rate, the contribution of $\hat\Sigma$ is much smaller than that of $\hat\Lambda$, since
  $\Sigma_i\approx \Lambda_i/100$.}
For the present analysis, the set of waveforms considered are the following:
\begin{itemize}
\item{$h_0$: 6PN order TaylorF2 waveform template bank, with zero spin-tidal 
contributions and setting to zero the tidal-tail contribution $\hat K=0$;}
\item{$h_{\chi}$: GW signal described by a $6.5$PN order TaylorF2 waveform accounting for irrotational fluids, with
  nonvanishing spins and $\hat K=0$.}
\end{itemize}
Note that in both cases we are not considering the tail-tidal term, imposing $\hat K=0$. The reason is that the
tail-tidal term limits the convergence domain of the $6.5$PN order term to frequencies $f\lesssim 100
\,\mathrm{Hz}$. Beyond these frequencies, including the $6.5$PN order tail contribution makes the accuracy of the PN
series worse than that retaining only terms up to $6$PN order (see \autoref{fig:65PNDets} below).  We discuss this issue
in detail in~\autoref{sec:trunc}.

Also note that, since the spin-tidal terms are linear in the spin and since we are neglecting quadratic and higher-spin
terms in the point-particle phase, the entire PN phase is symmetric under spin inversion, $\chi_i\to -\chi_i$.

\begin{figure}[!htb]
\includegraphics[width=\columnwidth]{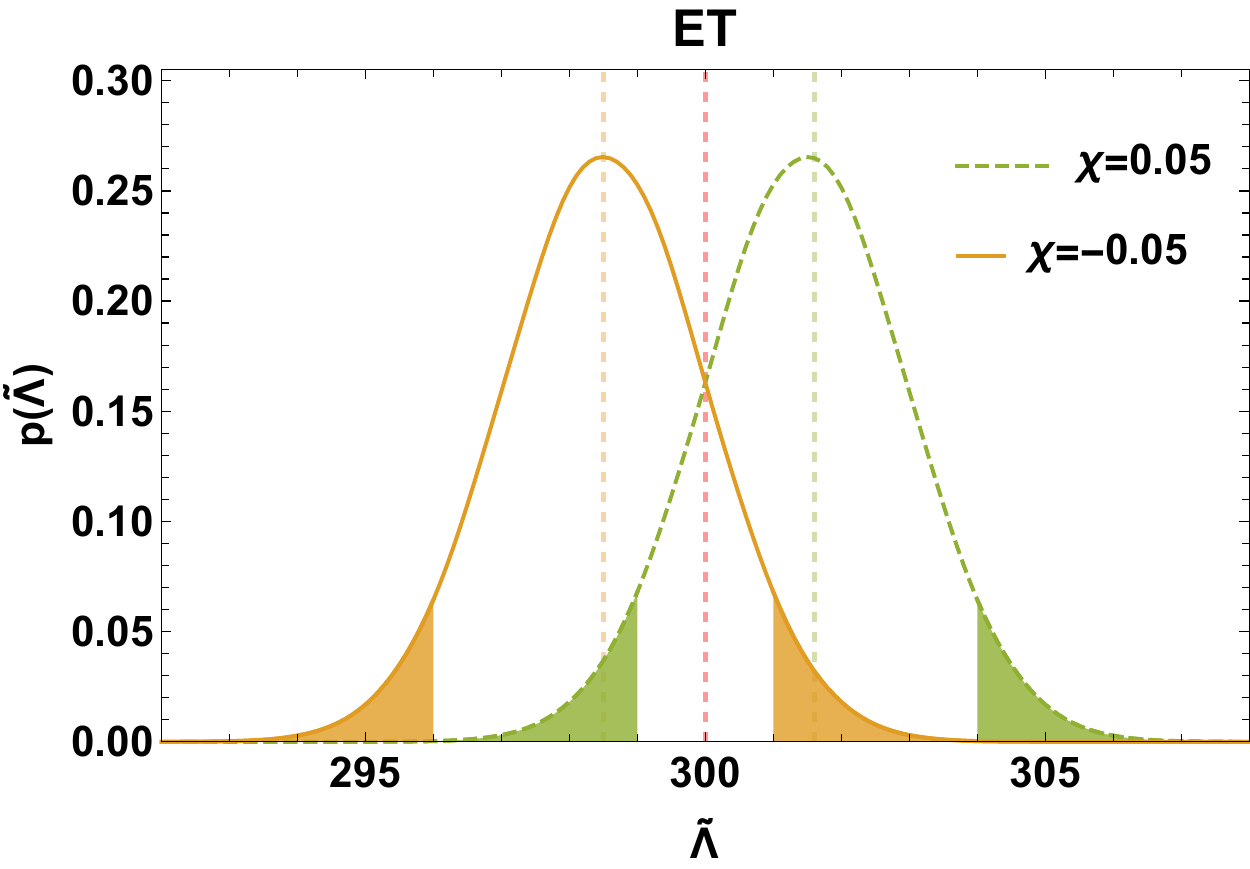}
 \caption{Probability distributions obtained in our \textit{standard} scenario for a binary spinning at $\chi=0.05$
   (dashed green line) and $\chi=-0.05$ (orange line), assuming irrotational fluids, with the \ET noise sensitivity curve. The
   vertical dashed lines define the best-likelihood values while the solid areas define the $90\%$ credible
   intervals. The red dashed vertical line defines the injected value $\tilde\Lambda_0$.
 \label{fig:65PNET300}
 }
 \end{figure}

In \autoref{fig:65PNET300} we show the effect of the spin-tidal corrections with respect to the standard 6PN order
approximant for the \ET noise sensitivity curve. The injected parameters are consistent with an equal-mass binary with
total mass $M=2.8M_\odot$, equal spins $\chi_1=\chi_2=0.05$, and $\tilde\Lambda_0=300$, thus setting a conservative
(\textit{standard}) scenario where the spins are compatible with current astrophysical observations.  Note that the
probability distributions match almost perfectly the nonspinning predictions described by $h_0$ though the spin-tidal
effects tend to induce a minimal shift on $p(\tilde\Lambda)$ that depends on the sign of the spin. The impact on the
recovery of $\tilde\Lambda$ is small, not producing a bias larger than $1\%$ with respect to $\tilde\Lambda_0$. The sign
of the offset tends to overestimate and underestimate $\tilde\Lambda$ for positive (dashed  green line) and negative
(orange line) spins, respectively. This can be explained intuitively by observing the relation between the electric $6.5$PN
order spin-tidal coefficient $\hat\Lambda$ and the $6$PN order $\tilde \Sigma$ one from \autoref{fig:6PNall}. We observe
that for $\chi_1=\chi_2=\pm0.05$ the corrections induced by these terms produce similar corrections to the orbital
phase, where the role of positive spins would be similar to that of static fluids while negative spins would affect
similarly to irrotational fluids. Moreover, notice the similarity between the offsets obtained in Figs.
\ref{fig:6PN800ET} and \ref{fig:65PNET300}, with the corrections on the former being larger due to the larger
$\tilde\Lambda_0$ considered. Thus, we see that the linear dependence of the orbital phase on each of the tidal
contributions induces a similar linear behavior on the bias produced when comparing different approximants. Thus,
\autoref{fig:65PNET300} shows that the spin-tidal coefficients for an event fully compatible with GW170817 are
negligible even when assuming a detection with \ET. This is also in agreement with the results obtained by full
hydrodynamical NR simulations of BNS systems~\cite{Dietrich:2017aum}, where the spin-tidal effects do not show any
significant contribution to the orbital phase for spins as high as $\chi \sim 0.15$.
\begin{figure}[!htb]
\includegraphics[width=\columnwidth]{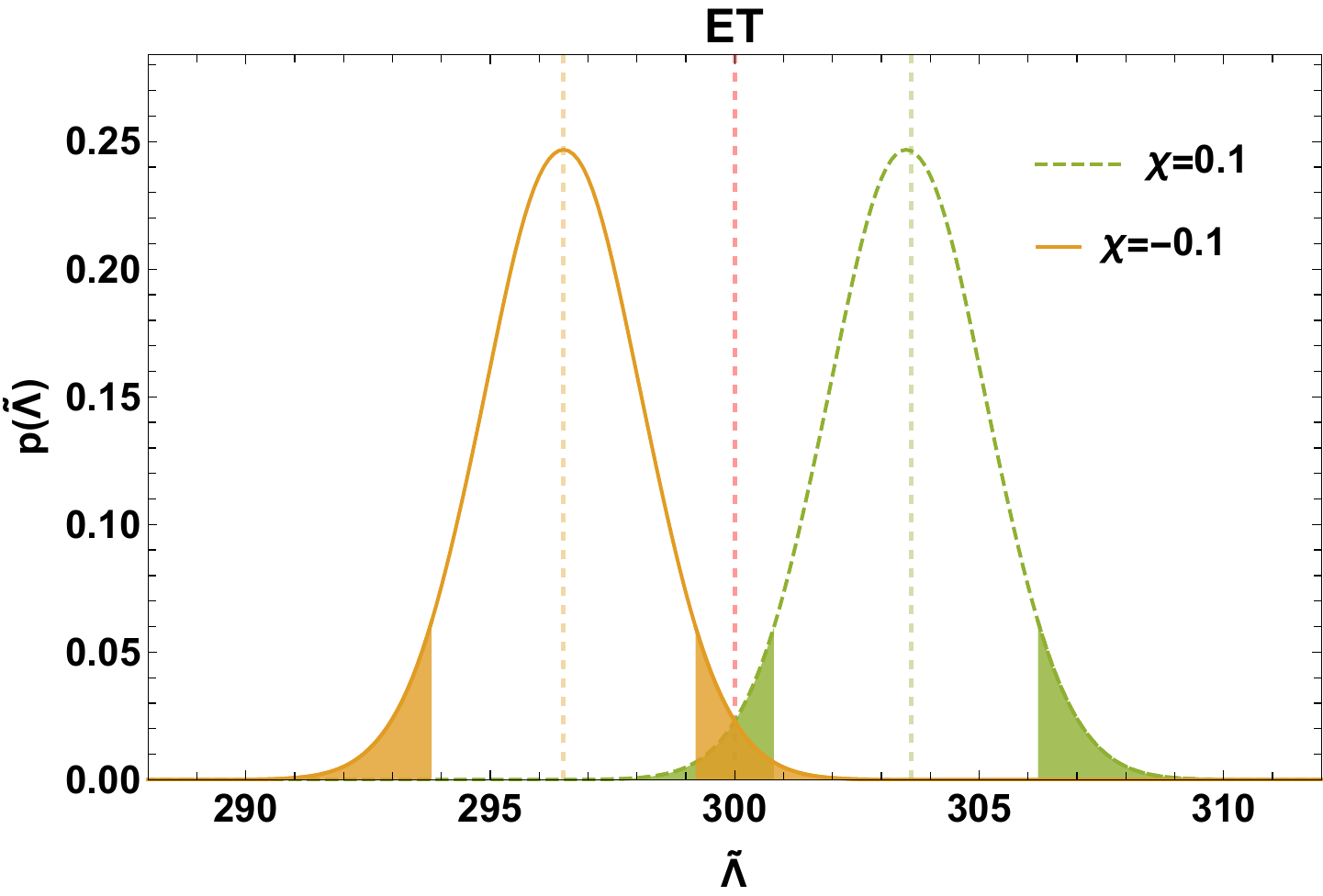}
\includegraphics[width=\columnwidth]{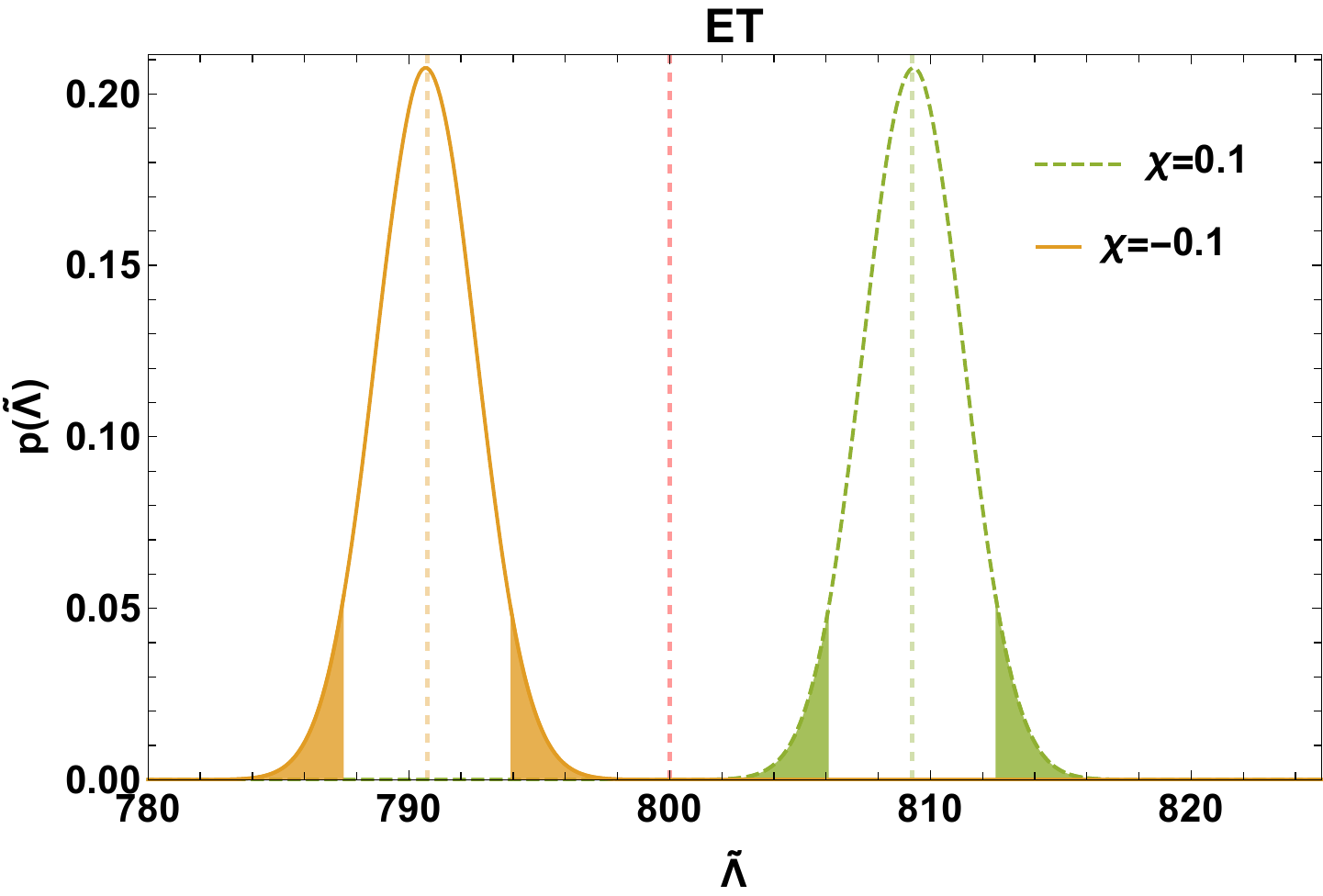}
 \caption{Probability distributions obtained for two BNS spinning at \comment{${\chi=0.1}$} (dashed green line) and  \comment{${\chi=-0.1}$}
   (orange line), assuming irrotational fluids, with the \ET noise sensitivity curve but now considering the case of
   observing GW170817 with an optimal orientation and ${\tilde{\Lambda}=\left\lbrace 300 ,800\right\rbrace}$ (top and bottom panels). The vertical dashed lines define the best-likelihood values while the
   solid areas define the $90\%$ credible intervals. The red dashed vertical line defines the injected value
   $\tilde\Lambda_0$.
 \label{fig:65PNET800}
 \vspace*{-\baselineskip}
 }
 \end{figure}
The picture slightly improves when we compute the deviations in our \textit{optimistic} scenario, that is, \comment{increasing
${\rho}$ by a factor of $1.15$} and setting ${\tilde\Lambda_0=\left\lbrace 300, 800\right\rbrace}$, but now also considering spin rates as high as
\comment{${\chi_1=\chi_2=\pm 0.1}$} in order to maximize the spin-tidal effects. The results for this case are shown in
\autoref{fig:65PNET800}. For ${\tilde\Lambda=300}$ the best-likelihood values for $\tilde\Lambda$ for both aligned (orange line) and antialigned
(dashed green line) binaries lie approximately on the tails of the $90\%$ credible intervals delimited by the solid areas,
thus being the spin-tidal waveforms $h_{\chi}$ marginally distinguishable from the template $h_0$. The peak offsets go in the
same direction as in \autoref{fig:65PNET300} but now largely increased because the SNR, the electric TLN
$\tilde\Lambda$, and the spins $\chi_{1,2}$ for the simulated events $h_{\chi}$ are a factor of $\sim 2.7$, $2$, and
\comment{$55$}  times larger, respectively (see Appendix \ref{sec:appdx}). \comment{In the pure optimistic scenario, that is, ${\tilde{\Lambda}=800}$, we observe  a much larger offset with respect to the small $\tilde{\Lambda}$ case of \comment{$6\sigma$} and thus clearly placing the injected value outside the $90\%$ credible intervals. Therefore, a spin-tidal model would produce a distribution substantially different from the pure 6PN models being its effects relevant for parameter estimation.}
\begin{figure}[!htb]
 \includegraphics[width=\columnwidth]{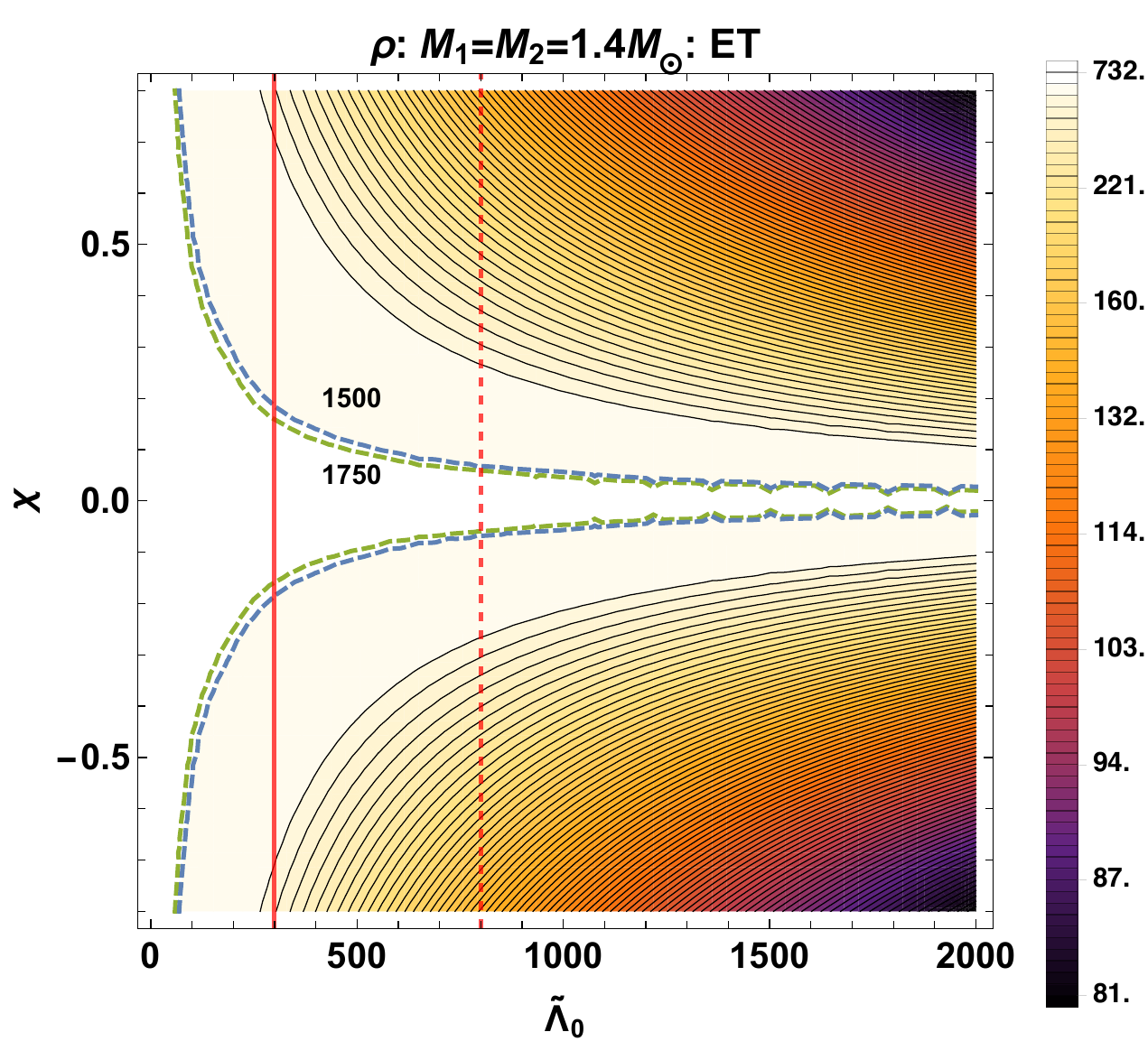}
 \caption{Estimation of the \snr required to distinguish the effects of the 
spin-tidal terms considering the \ET noise
   sensitivity curve. The vertical red grid lines fix the tidal deformabilities consistent with the median and $90\%$
   upper limits provided by LIGO-Virgo~\cite{TheLIGOScientific:2017qsa,Abbott:2018wiz} respectively. The blue and green
   contour lines correspond to the \snr of our \textit{standard} and \textit{optimistic} scenarios. 
 \label{fig:Match}
 \vspace*{-\baselineskip}
 }
 \end{figure}
 
Finally, we also provide an estimate of the minimum values for the triplet ${\chi-\rho-\tilde\Lambda_0}$ (with
$\chi_1=\chi_2=\chi$) required to distinguish the effects of the 
spin-tidal terms for a GW170817-like event detected with \ET. To do so, we have 
computed the match of a
$h_0(\tilde\Lambda_0,\chi)$ against $h_{\chi}(\tilde\Lambda_0,\chi)$, for $\tilde\Lambda_0\in[0,2000]$ and
$\chi\in[-0.79,0.79]$. The results of the match are translated to $\rho$ through \autoref{eq:ind} for a ${D=6}$
parameter space, where we require one to estimate all the parameters at $90\%$ credible level, i.e., ${n=1.64}$. The results
of this analysis are shown in \autoref{fig:Match}. The contour lines represent the minimum \snr needed to observe some
characteristic combination of $\tilde\Lambda_0$ and ${\chi_1=\chi_2=\chi}$. The solid and dashed vertical grid lines
${\tilde\Lambda_0=\left\lbrace 300,800\right\rbrace}$ set the median and $90\%$ upper limit provided by
~\cite{TheLIGOScientific:2017qsa,Abbott:2018wiz}, respectively. Then, the intersection of ${\tilde\Lambda_0=800}$ with
the \comment{${\rho=\left\lbrace 1500,1750\right\rbrace}$ contours} shows that the minimum spin required to distinguish the
spin-tidal effects from a $h_0$ template at the $90\%$ level is \comment{${\chi \sim \pm 0.07}$},
respectively.  Notice that the intersection of the $\tilde\Lambda_0=800$ line with the green contour line \comment{${(\rho=1750)}$} corresponds to the particular case shown in \autoref{fig:65PNET800}.
Moreover, from \autoref{fig:Match}, we see that larger spins are required to attain the same SNR as $\tilde\Lambda_0$
decreases. In particular, for $\tilde\Lambda=300$ and \comment{${\rho=1750}$} the intersection occurs at \comment{${\chi \sim \pm 0.15}$}.
Therefore, spin-tidal couplings are only expected to affect significantly the signal for putative \comment{optimally oriented} BNS events, observed with third-generation detectors, and for \comment{moderately} large spins. On the other hand, the
calibration of these effects on current waveform templates would have a non-negligible impact only if high-spin binaries
(with \comment{$\chi_i\gtrsim0.1$}) evolve and merge in our local universe.

Finally, we note that we have also estimated the one-dimensional probability distributions on $p(\tilde\Lambda)$ by
running a six-dimensional Markov chain Monte Carlo algorithm on \autoref{eq:probFMmr}, where $p(\tilde\Lambda)$ is
obtained by marginalization.  By doing so, we did not observe any relevant differences with respect to the distributions
$p(\tilde\Lambda)$ obtained in this section, thus suggesting that the correlations between the physical parameters do
not affect our results in such high SNR scenarios.

\section{Truncation effects on high-PN order TaylorF2 terms}\label{sec:trunc}
PN models approximate the orbital dynamics by a power-series expansion of the equations in terms of the parameter
$x=v^2/c^2 = (G\omega M)^{2/3}/c^2\ll1$. However, in the high-frequency regime the optimal truncation order may be
limited by the convergence properties of the PN series. This has been extensively studied in the case of binary black
holes, where the expansion above $3$PN order is shown to break down at relatively low frequencies
~\cite{Cutler:1992tc,Simone:1996db,Yunes:2008tw,Zhang:2011vha}. In this section we study the properties of the tidal
part of the PN series as an \emph{asymptotic} series~\cite{bender2013advanced}. Formally, a power series is said to be
asymptotic to a function $f(x)$ as $x \to x_0$ if for each $N$
\begin{equation}
\label{eq:asymptotic}
f(x)- \sum_{n=0}^N a_n (x-x_0)^n \ll a_N (x-x_0)^N \,.
\end{equation}
This equation states that, to satisfy the asymptotic condition near some point $x_0$, the difference between a function
and the $N$-truncated sum of the series should be much smaller than the last term kept in the expansion.
If the series is divergent [or it is not converging to $f(x)$], for each given point $x$ there is a maximum order
$N=N(x)$ for which the match between the function and the series truncated up to that order is the optimal one, which
means that including higher-order terms will decrease the accuracy of the approximation. For the PN case
\autoref{eq:asymptotic} reduces to
\begin{equation}
\label{eq:convcrit}
\psi(x)- \sum_{n=0}^N a_{n/2} \, x^{n/2} \ll a_{N/2} \, x^{N/2}\,,
\end{equation}
where $2n$ is the PN order [see \autoref{PHASE}] and the function $\psi(x)$ is the exact (but unknown) full
gravitational waveform phase for the binary under consideration. Likewise, the range of validity of the truncated
expansion at some fixed PN order can be limited to some maximum point $x=x_{max}$, above which \autoref{eq:convcrit} is
no longer satisfied. The exact value of $x_{max}$ will in general vary across the parameter space (component masses,
spins, etc.) though for BNS systems we expect this variation to be smaller than for binary black holes due to the
relative smallness of the parameter space.

Here we study the asymptotic behavior of the PN tidal terms in the case of a nonspinning, equal-mass BNS with vanishing
magnetic TLNs. The latter approximation should not affect our analysis since, as shown in the previous sections,
magnetic TLNs give a negligible contribution to the GW phase. Then, the tidal approximants can be considered valid for
parameter estimation studies if and only if all these terms satisfy \autoref{eq:convcrit}, across the full parameter
space and along the full frequency regime of ground-based detectors.

To illustrate this, we assume the NR-calibrated model IMRPhenomD-NRTidal~\cite{Dietrich:2017aum} as the true underlying tidal
part of GW phase $\psi(x)$ in \autoref{eq:convcrit}, recalibrating its coefficients to also recover the $6.5$PN order
tail coefficient $\hat{K}$ in the low-frequency limit. Doing so we ensure that the model by itself represents
effectively the same phase evolution as the original one but achieving a better match with the PN solutions at
intermediate frequencies. Then, we check whether \autoref{eq:convcrit} is satisfied for all the truncated expansions of
the TaylorF2 approximant previously considered. For this analysis we adopt the physical parameters of GW170817; in
particular we consider nonspinning binaries.
 \begin{figure}[t]
\includegraphics[width=\columnwidth]{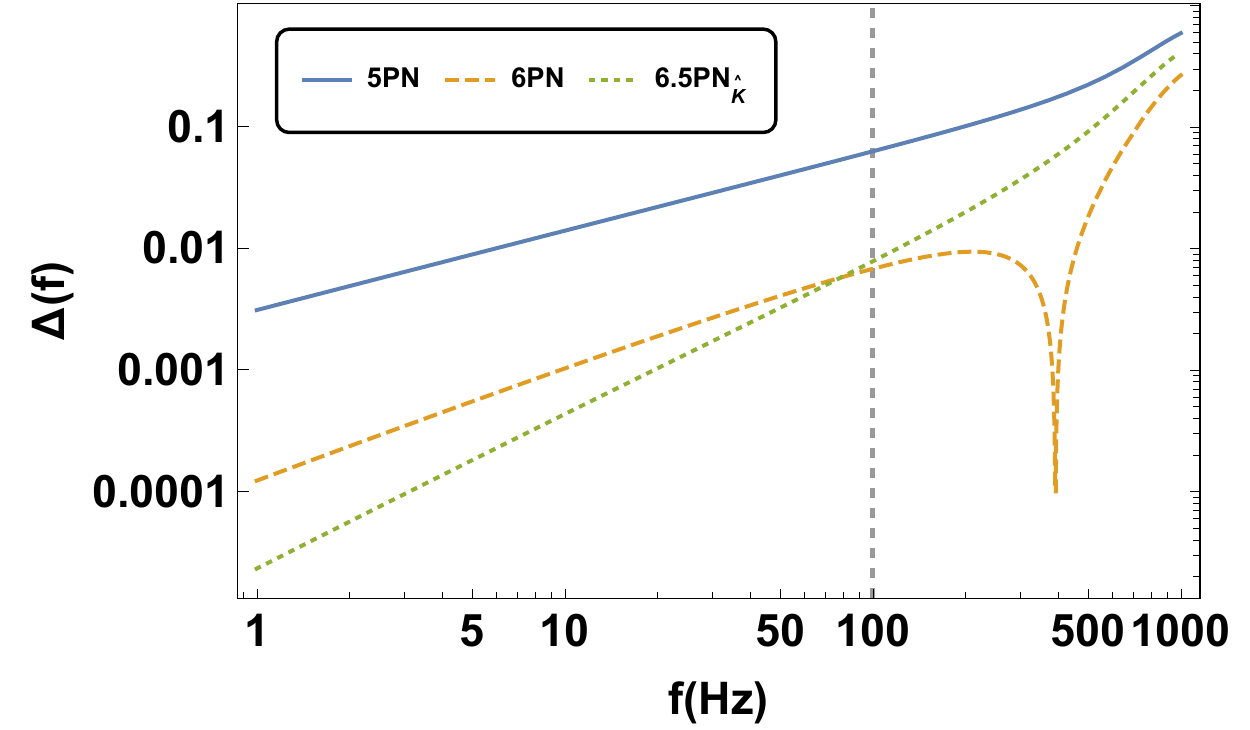}
 \caption{Comparison between the TaylorF2 and PhenomD approximants following the criterion defined in
   \autoref{eq:convcrit}. We show $\Delta(f)$ [cf. \autoref{eq:delta}] as a function of the frequency for the same set
   of tidal corrections.
The vertical dashed line at $f = 100 \, \mathrm{Hz}$ fixes the approximate frequency where the discrepancies $\Delta(f)$
for the 6PN order and the $6.5{\rm PN_{\hat{K}}}$ order tidal-tail term cross each other for an equal-mass $2.8M_\odot$
BNS system.
\label{fig:65PNDets}
 }
 \end{figure}
In  \autoref{fig:65PNDets} we show the quantity
\begin{equation}
 \Delta(x)=\left|\psi(x)- \sum_{n=0}^N a_{n/2} \, x^{n/2} \right|\,
\label{eq:delta}
\end{equation}
[i.e.,the left-hand side of \autoref{eq:convcrit}] for each PN tidal order, namely
$\left\lbrace 5{\rm PN}, 6{\rm PN}, 6.5{\rm PN_{\hat{K}}}\right\rbrace$, where $6.5{\rm PN_{\hat{K}}}$ indicates the
tidal-tail term entering at 6.5PN order (i.e., neglecting the spin-tidal part entering at the same order). The
asymptotic condition in \autoref{eq:delta} is satisfied by requiring only that including higher-order PN terms increases
the agreement between the series and the full IMRPhenomD-NRTidal model, i.e., that $\Delta(x)$ decreases as more terms are added
to the series. From \autoref{fig:65PNDets} we notice that for frequencies approximately below $100 \, \mathrm{Hz}$ this
is indeed the case. However, this is not true anymore above
$f\sim100 \, \mathrm{Hz}$ which roughly corresponds to the crossing between the $6$PN order curve and the $6.5{\rm
  PN_{\hat{K}}}$ order curve in \autoref{fig:65PNDets}.
In other words, including the $6.5{\rm PN_{\hat{K}}}$ order tidal-tail term makes the difference between the IMRPhenomD-NRTidal
model and the PN series larger than that obtained retaining only terms up to $6$PN order. Based on this, we can assert
that the addition of the $6.5{\rm PN_{\hat{K}}}$ order tail-tidal term will only improve the PN approximants in the low
frequency regime. On the other hand, at frequencies $f \gtrsim 100 \, \mathrm{Hz}$, the inclusion of this term decreases
the agreement between the TaylorF2 approximant and the IMRPhenomD-NRTidal model. Considering that current ground-based detectors
collect most of the \snr around these frequencies, the inclusion of such a term would be magnified, producing a negative
impact on parameter-estimation analyses.

It is also worth noting that, since the IMRPhenomD-NRTidal model does not account for the spin-tidal interactions, we could not
extend this analysis to the new $6.5{\rm PN_{\hat{\Lambda}}}$ order spin-tidal contributions $\hat\Lambda$. This has
prevented us from determining whether there exists a maximum frequency $x_{max}$ for the $\hat\Lambda$ coefficient for which
\autoref{eq:convcrit} is unfulfilled, thus not ensuring the correctness of such terms up to $f_{max}=2048$. However, as
it is shown in \autoref{fig:6PNall} and in agreement with the results of \autoref{sec:res}, the order of magnitude of
these terms is expected to be a factor ${\sim 8}$ smaller than the tidal-tail term for moderately high spins
\comment{${\chi=0.1}$}. Therefore,  the corrections being so small suggest that there are no such divergences for the $6.5{\rm
  PN_{\hat{\Lambda}}}$ order term. For this reason, we set ${f_{max}= 2048}$.

We quantify the impact of different PN tidal terms by applying the analysis described in~\autoref{sec:stat}. In this
case, we use as the reference model the full IMRPhenomD-NRTidal model matched against the tidal part of the TaylorF2 approximant
truncated to $\left\lbrace 5{\rm PN}, 6{\rm PN}, 6.5{\rm PN_{\hat{K}}}, 6.5{\rm PN_{\hat{\Lambda}}} \right\rbrace$
order, respectively, in the frequency range $f\in(23,2048)$~{\rm Hz}, consistent with the range used
in~\cite{Abbott:2018wiz}. The $6.5{\rm PN_{\hat{K}}}$ and $ 6.5{\rm PN_{\hat{\Lambda}}}$ terms refer to two separate
$6.5$PN order models, where in the former we  include only the tidal-tail term ${\hat K}$ while in the latter we set
${\hat K=0}$ but accounting for the spin-tail coefficient ${\hat \Lambda}$ with ${\chi=0.2}$.

\begin{figure}[!htb]
\includegraphics[width=\columnwidth]{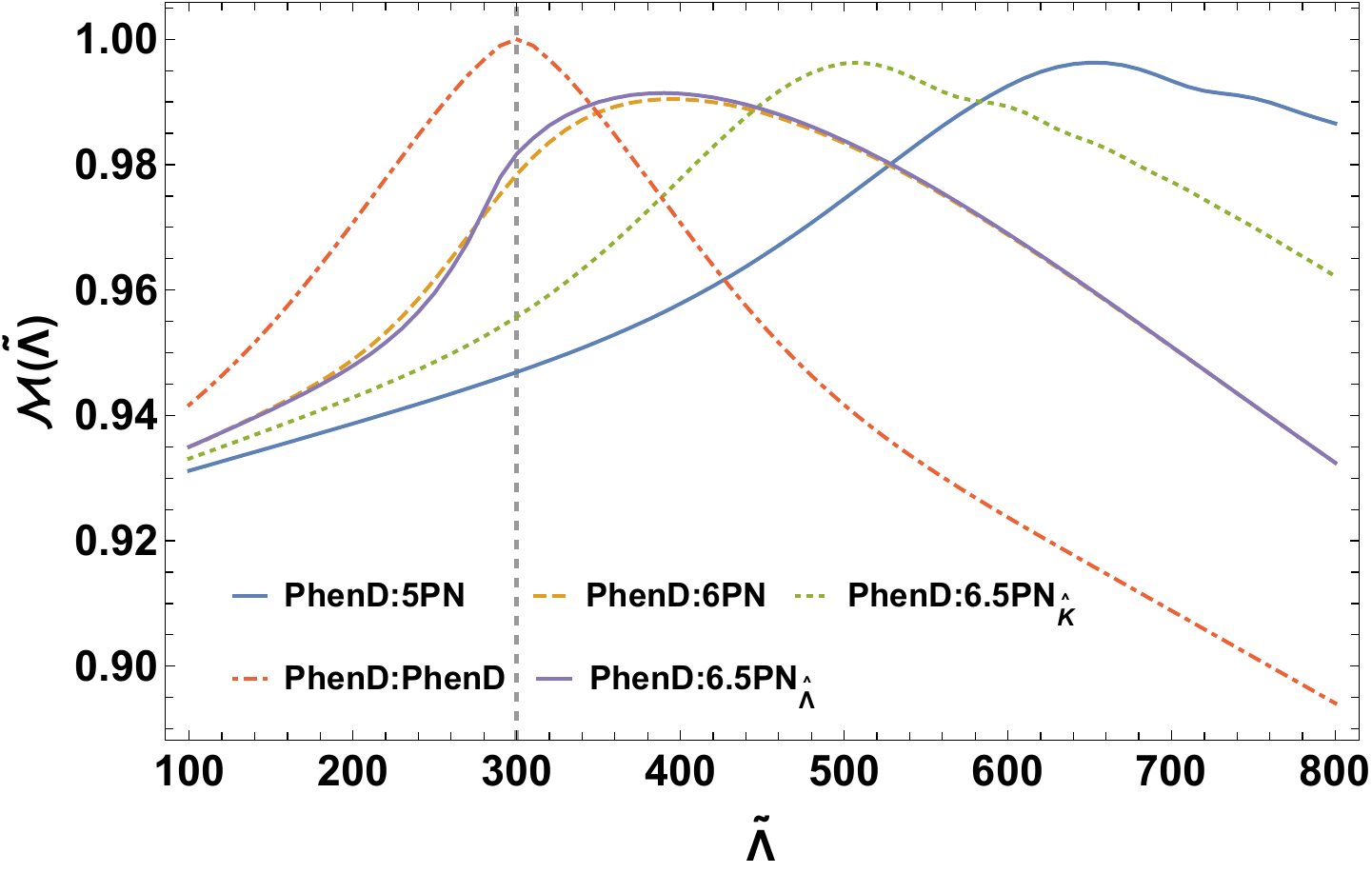}
 \caption{Match as in \autoref{eq:match} obtained by comparing the PhenomD model and the TaylorF2 approximant truncated
   at $\left\lbrace 5{\rm PN}, 6{\rm PN}, 6.5{\rm PN_{\hat{K}}}, 6.5{\rm PN_{\hat{\Lambda}}}\right\rbrace$ order.  The
   $6.5{\rm PN_{\hat{K}}}$ and $ 6.5{\rm PN_{\hat{\Lambda}}}$ terms refer to the tidal-tail and spin-tail coefficients
   respectively with ${\chi=0.2}$. The dashed vertical line determines the injected tidal deformability for an
   equal-mass binary with total mass $M=2.8M_\odot$.
 \label{fig:mmPN}
 }
 \end{figure} 
 
In \autoref{fig:mmPN} we show the effect of adding sequentially new PN terms to the TaylorF2 approximant. The
first term considered in this analysis is the leading-order $5$PN term, which provides an estimated value of the
weighted tidal deformability $\tilde\Lambda$ remarkably shifted about ${130\%}$ from the injected one (blue curve). The
offset is significantly reduced by adding the $6$PN order term (orange curve) to the TaylorF2 approximant, which now
makes the reconstructed $\tilde\Lambda$ a value ${30\%}$ closer to the true one. This suggests that the $6$PN order term
still accounts for a non-negligible contribution to the tidal phase; thus it cannot be omitted for parameter
estimation. Indeed, the TaylorF2 approximant up to $6$PN order has been used to provide the first estimates to the
leading-order tidal deformability in \cite{TheLIGOScientific:2017qsa}.  Interestingly, our analysis suggests that
TaylorF2 waveforms always provide upper bounds on $\tilde\Lambda$ larger than those provided by NR-calibrated
waveforms~\cite{TheLIGOScientific:2017qsa}.
Furthermore, the addition of the next-to-next-to-leading-order $6.5{\rm PN_{\hat{\Lambda}}}$ spin-tidal term (purple line)
with ${\chi=0.2}$ just produces a minimal deviation with respect to the $6{\rm PN}$ next-to-leading-order term due to
the smallness of the spin-tidal interaction, which is in agreement with the results of \autoref{sec:res}. On the
contrary, the tail-tidal term (green curve) does not improve the agreement between the two waveform models but actually
increases the systematic error. We believe that this is due to the fact that we are using such term in a regime where
the PN series is poorly convergent, so that such term is no longer a good approximation to the true GW phase.  Including
(currently unknown\footnote{Here we are not considering the EOB tidal model, in which some terms have been derived up to
  $7.5$PN order~\cite{Damour:2012yf}.})  higher-order PN terms could in principle correct this pathological behavior.

\section{Conclusion}\label{sec:conclusions}
\vspace{-\baselineskip} We have estimated the impact of the $6$PN order magnetic TLNs terms described
in~\cite{Yagi:2013sva,Banihashemi:2018xfb,Abdelsalhin:2018reg} and the new $6.5$PN order spin-tidal corrections computed
by~\cite{Abdelsalhin:2018reg} on GW BNS events with physical parameters consistent with GW170807. We considered two
different scenarios: a \textit{standard} scenario in which we choose the physical parameters to be consistent with the
median estimates provided by the LIGO-Virgo Collaboration \cite{Abbott:2018wiz,TheLIGOScientific:2017qsa}
and~\cite{De:2018uhw}, and a more \textit{optimistic} scenario based on a hypothetical optimally oriented GW170807-like
event (\comment{thus increasing the \snr to ${\rho=1750}$ or to that compatible with the optimal observation of GW170817 under ET}) and where the leading-order tidal deformability is taken to be
$\tilde\Lambda=800$, i.e., consistent with the LIGO-Virgo upper limits. Given the high \snr produced in the two
scenarios, we have taken advantage of the FIM formalism to compute the parameter bias induced by neglecting the magnetic
TLNs and the spin-tidal terms on our waveform approximants. We provide estimates for both scenarios and for the \aLIGOe
and \ET noise sensitivity curves.

We found that the internal dynamics induced on the NS fluid and encoded in the magnetic TLNs \comment{could be observed} with third-generation GW detectors such as \ET for our optimistic scenario. The effects of the magnetic TLNs, for both static and
irrotational fluids, do not affect by more than \comment{${\sim 5 \%}$} the estimate of $\tilde\Lambda$ when they are not included
in current waveform template banks. This is explained by the fact that the magnetic TLNs are roughly a factor of $100$
smaller than the corresponding electric TLNs, while they enter at the next-to-leading PN order relative to the principal
$\tilde\Lambda$. \comment{This is consistent with \cite{Bernuzzi:2014owa}.} 

We find a slightly more optimistic scenario regarding the recently computed 6.5PN order spin-tidal couplings. In this
case we have quantified for the first time the impact of these terms by means of the bias produced on the measurement of
the tidal deformability $\tilde\Lambda $ that arises from neglecting these terms in our waveform approximants. We find
that for a GW170817-like BNS event detected by the ET in the optimal orientation, spin-tidal effects can be negligible
unless the spins are at least \comment{${|\chi_{1,2}| \sim 0.07}$} for ${\tilde\Lambda \sim 800}$ (which is the LIGO-Virgo upper
limit on $\tilde\Lambda$) and \comment{${|\chi_{1,2}| \sim 0.15}$} for the more conservative value ${\tilde\Lambda \sim
  300}$. Therefore, these effects could be relevant for BNS waveform approximants only in the unlikely case that
spinning binaries with \comment{${\chi_{1,2}\sim 0.1}$} merge in our local universe. However, considering the current accuracy
of the NR codes, we do not expect that the minimal variations produced by spin-tidal couplings can be separated from the
numerical noise \cite{Dietrich:2017aum}.
We focused on the planned ET detector~\cite{Hild:2009ns,Sathyaprakash:2011bh,Sathyaprakash:2012jk}, but similar results
are expected for other third-generation designs, such as Cosmic Explorer~\cite{Evans:2016mbw,Essick:2017wyl}. In
particular, since the minimum sensitivity of the latter is a factor of a few better than ET, we expect that the effect
of spin-tidal coupling should be slightly easier to detect.

Finally, we have studied the convergence properties of the high-PN order tidal terms. This is relevant for any study
willing to add higher than 6PN order terms to TaylorF2 approximants. We found that though these terms contribute to
increase the accuracy at the very low frequencies (${f \ll 100\,{\rm Hz}}$), they do not satisfy the convergence
properties at ${f \sim 100\,{\rm Hz}}$.
Given that, for BNS events, most of the \snr is collected at frequencies that surround this value, the inclusion of such
higher-order terms in the waveform models may lead to large systematic errors and to a significant bias on the parameter
estimation. This explains why TaylorF2 approximants restricted to 6PN order produce results more compatible with the
NR-calibrated models than the extended 6.5PN order tidal-tail model, since the 6PN order term satisfies the convergence
properties, whereas this is not the case of the tidal-tail term.
We could not extend this analysis to the 6.5PN order spin-tidal coefficients since there is no NR-calibrated approximant
accounting for these terms. Alternatively, a different resummation of the PN terms as in \cite{Nagar:2018zoe} could
correct the pathological behavior of the series.

Future work will focus on the inclusion of the RTLN terms computed in Ref.~\cite{Abdelsalhin:2018reg}, although this
will have to wait until the conceptual problem related to the inclusion of the RTLNs in a Lagrangian formulation is
solved.
Another extension of our work is related to the analysis of systems for which the spin-tidal effects are expected to be
larger, for instance in GW searches for exotic compact objects based on tidal
effects~\cite{Cardoso:2017cfl,Sennett:2017etc,Maselli:2017cmm}. There is no reason to expect that such objects should be
slowly spinning (this is particularly true for supermassive objects in the LISA band, whose spin might grow through
accretion or through subsequent mergers during the galaxy evolution). We expect that the inclusion of the spin-tidal
couplings computed here will improve previous analysis~\cite{Maselli:2017cmm}.
\section*{Acknowledgments}

\vspace{-0.5\baselineskip}
We  thank Evan Hall for making the computations of the Einstein 
Telescope geometrical factors available and Ilya Mandel for useful clarifications about 
our computations.
We acknowledge support from the Amaldi Research Center funded by the
MIUR program ``Dipartimento di Eccellenza" (CUP: B81I1800101). P.P. acknowledges financial support provided under the European Union's H2020 
ERC, Starting Grant Agreement No.~DarkGRA--757480, and the kind hospitality of 
the Universitat de les Illes Balears, where part of this work was done.
The authors acknowledge networking support by the COST Action 
CA16104. 
This project has received funding from the European Union's Horizon 2020 research and innovation program under the
Marie Sklodowska-Curie Grant Agreement No 690904.

\appendix
\section{APPROXIMATED CONVERSION OF GW170817  TO ET}
\label{sec:appdx}
A GW strain $h(t)$ detected by some particular GW observatory in its own coordinate frame 
has the following form,
\begin{align}
\label{eq:hproj1}
h_{D}(t,\vec{\gamma},\iota,\theta,\phi,\psi, \zeta)= D_A^{ij}(\theta,\phi,\psi, \zeta)h_{ij}(t,\vec{\gamma},\iota) \,,
\end{align}
where $D_A^{ij}$ is the so-called \textit{detector tensor} of the detector $A$ and  $h_{ij}$  accounts for the GW strain tensor. Then, \autoref{eq:hproj1} is nothing other than the projection of the strain tensor $h_{ij}$ defined in the source coordinate frame (where $\vec{\gamma}$ are the physical parameters and $\iota$ the source inclination) to the detector frame according to its sky location ${\left\lbrace\theta,\phi\right\rbrace}$,  polarization angle $\psi$, and the angle between the detector arms $\zeta$. In other words, 
it gives the amount of signal traveling in the direction perpendicular to the detector plane and with polarization angle $\psi$. For instance, an interferometer with a pair of arms forming an angle  $\zeta$, the detector tensor $D_A^{ij}$ reads
\begin{equation}
D_A^{ij}=\frac{1}{2}
\begin{bmatrix}
\label{eq:Dtop1}
F_+(\theta,\phi,\psi) \sin^2{\zeta} & - F_\times(\theta,\phi,\psi) \cos{\zeta}\sin{\zeta}  \\
- F_\times(\theta,\phi,\psi) \cos{\zeta}\sin{\zeta}     & -F_+(\theta,\phi,\psi) \sin^2{\zeta}
\end{bmatrix}\,,
\end{equation}
where $F_{+,\times}(\theta,\phi,\psi, \zeta)$ are the so-called detector antenna patterns. In parallel, the strain tensor $h_{ij}$ of an elliptically polarized GW traveling perpendicular to the detector frame is
\begin{equation}
h_{ij}=h_0(t,\vec{\gamma}) 
\begin{bmatrix}
\label{eq:hij}
 \frac{1+\cos^2{\iota}}{2} &  i \cos{\iota} \\
i  \cos{\iota}    & - \frac{1+\cos^2{\iota}}{2}
\end{bmatrix}\,,
\end{equation}
consistent with the $h_+$ (diagonal) and $h_\times$ (antidiagonal) polarizations and 
noting that $\iota=0$ represents the case of the optimally oriented (circularly polarized) source  considered in 
this work. For an L-shaped detector such as the LIGO-Virgo observatories ($\zeta=90º$) it 
is easy to show that combining equations \ref{eq:hproj1}, \ref{eq:Dtop1} and \ref{eq:hij}
we get the usual expression for $h_D$,
\begin{align}
\label{eq:rhoL}
h_D(t,\vec{\gamma},\iota,\theta,\phi,\psi, \zeta)=\big(\frac{1+\cos^2{\iota}}{2}\big) F_+(\theta,\phi,\psi) h_+(t,\vec{\gamma}, i ) \,,
\end{align}
which simplifies to
\begin{align}
\label{eq:hpol}
h_D(t,\vec{\gamma},\iota,\theta,\phi,\psi, \zeta)=\big(\frac{1+\cos^2{\iota}}{2} \big) h_0(t,\vec{\gamma} ) \,,
\end{align}
for the particular case of ${\theta=\phi=\psi=0}$.

On the other hand, the \snr collected by a network of (uncorrelated)  detectors is given by
\begin{equation}
\label{eq:rhotot}
\rho=\sqrt{\sum_D \rho_D^2}\,,
\end{equation}
where $\rho_D$ accounts for the \snr observed by a single detector as described by \autoref{eq:inner}. In the particular case of GW170817, its position in the sky implied that almost all the \snr was collected by the Hanford and Livingstone observatories \cite{TheLIGOScientific:2017qsa} also with similar magnitudes. In this scenario, \autoref{eq:rhotot} can be approximated  by
\begin{equation}
\label{eq:rhoL2}
\rho_{HL}=\sqrt{\rho_{H}^2 +\rho_{L}^2} \approx \sqrt{2} \,\rho_{H} \approx 
\begin{cases}
    \rho_{HL}^{st}=32.4 \\
    \rho_{HL}^{opt}=32.4\big(\frac{2}{1+\cos^2(30º)}\big)
\end{cases}
\end{equation}
where $\rho_{H,L}$ are the individual Hanford and Livingstone \snr's and  the first 
line consistent with our standard scenario $\rho_{HL}^{st}$ while in the second one $\rho_{HL}^{opt}$ reproduces the same 
LIGO-Virgo event if $\iota$ would have been optimal. Alternatively, current prospects 
concerning the design of the ET geometry anticipate the construction of a detector formed 
by joining three separate interferometers in a triangular shape, that is, with $\zeta=60º$ 
\cite{Hall,Sathyaprakash:2012jk} and with the three responses described by 
$D_1^{ij},D_2^{ij}$ and $D_3^{ij}$ as
\begin{align}
\label{eq:Dtop}
D_{1,2}^{ij}=\frac{1}{2}
\begin{bmatrix}
\pm  \frac{3}{4} & - \frac{\sqrt{3}}{4}  \\
- \frac{\sqrt{3}}{4}    & \mp \frac{3}{4}
\end{bmatrix},\qquad 
D_{3}^{ij}=\frac{1}{2}
\begin{bmatrix}
0 &  \frac{\sqrt{3}}{2}  \\
 \frac{\sqrt{3}}{2}    & 0
\end{bmatrix}\
\,,
\end{align}
where we have also assumed ${\theta=\phi=\psi=0}$.  Then, using Eqs. \ref{eq:hproj1}, \ref{eq:rhotot},
and \ref{eq:Dtop}, the total \snr in ET will result from 
adding up the three individual contributions as,
\begin{equation}
\label{eq:rhoET}
\rho_{ET_\triangle} =\sqrt{\sum_{A=1,3} 4\mathcal{R}  \int_{f_{min}}^{f_{max}}\frac{|D_A^{ij}h_{ij}|^2}{S_n^{ET}(f)}df}  = f(\iota) \, \rho_{ET}\,
\end{equation}
and
 \begin{equation}
\label{eq:rhoET2}
f(\iota) = \frac{3}{2} \sqrt{1/8(1 + 6 \cos^2{\iota} + \cos^4{\iota})}\,,
\end{equation}
where $\rho_{ET}$ is the \snr computed for a single L-squared ET detector for an optimally oriented source and $f(\iota)$ is a geometrical factor equal to $3/2$ for $\iota=0$ \cite{Hall}. Then, by combining Eqs. \ref{eq:rhoL2} and \ref{eq:rhoET}, we get
 \begin{equation}
\label{eq:ET}
\rho_{ET_\triangle}=  f(\iota)\sqrt{ \frac{4\mathcal{R}\int_{f_{min}}^{f_{max}}\frac{\tilde{h}(f)\tilde{h}^*_T(f)}{S_{n}^{ET}(f)}
df}{4\mathcal{R}\int_{f_{min}}^{f_{max}}\frac{\tilde{h}(f)\tilde{h}^*_T(f)}{S_{n}^{H}(f)}
df}}\, \frac{\rho_{HL}^{opt}}{\sqrt{2}} \,,
\end{equation}
where $S_n^{ET,H}$ are the sensitivity curves taken from the literature for \ET and \aLIGOe, ${f_{min}=\left\lbrace 3,30\right\rbrace}$  and $f_{max}=2048 Hz$ for both detectors, respectively. Finally, we get the following conversion factors:
\begin{equation}
    \rho_{ET_\triangle}\approx 
 \rho_{HL}^{st} \begin{cases}
    45 &  \text{standard scenario}\\
    55             & \text{optimal scenario}
\end{cases}\,.
\end{equation}
\bibliography{./biblio.bib}
\end{document}